\documentclass[twocolumn,prd,aps,floatfix]{revtex4}

\usepackage{amsmath}
\usepackage{latexsym}
\usepackage[psamsfonts]{amssymb}
\usepackage{graphicx}
\usepackage{longtable}
\usepackage{epstopdf}
\usepackage{epsfig}
\usepackage{bm}
\usepackage{color}

\newcommand{\be}{\begin{equation}}
\newcommand{\ee}{\end{equation}}
\newcommand{\bea}{\begin{eqnarray}}
\newcommand{\eea}{\end{eqnarray}}
\newcommand{\bi}{\begin{itemize}}
\newcommand{\ei}{\end{itemize}}

\newcommand{\Pslash}{p \kern -2mm /}
\newcommand{\Btob}{B_1\rightarrow B_2}

\newcommand{\XtoS}{\Xi\rightarrow \Sigma}

\newcommand{\StoN}{\Sigma\rightarrow N}
\newcommand{\NtoS}{N\rightarrow \Sigma}

\newcommand{\LtoN}{\Lambda \rightarrow N}
\newcommand{\XtoL}{\Xi \rightarrow \Lambda}

\begin{document}

\title{
Continuum limit of hyperon vector coupling $f_1(0)$ \\
from 2+1 flavor domain wall QCD
}
\author{Shoichi Sasaki} 
\email[E-mail: ]{ssasaki@nucl.phys.tohoku.ac.jp}

\affiliation{Department of Physics, Tohoku University, Sendai 980-8578, Japan}

\date{\today}
\begin{abstract}
We determine the hyperon vector couplings $f_1(0)$ for $\Sigma^{-}\rightarrow nl^-\bar{\nu_l}$
and $\Xi^0\rightarrow\Sigma^{+}l^-\bar{\nu_l}$ semileptonic
decays in the continuum limit with (2+1)-flavors of dynamical domain-wall fermions,
using the Iwasaki gauge action at two different lattice spacings 
of $a= 0.114(2)$ and 0.086(2) fm.
A theoretical estimation of flavor $SU(3)$-breaking effect on 
the vector coupling is required to extract $V_{us}$ from the experimental 
rate of hyperon beta decays. We obtain the vector couplings $f_1(0)$
for $\StoN$ and $\XtoS$ beta-decays with an accuracy of less than one percent.
We then find that lattice results of $f_1(0)$ 
combined with the best estimate of $|V_{us}|$ 
with imposing Cabibbo-Kobayashi-Maskawa (CKM) unitarity are slightly deviated from 
the experimental result of $|V_{us}f_1(0)|$ for 
the $\StoN$ beta-decay. This discrepancy can be attributed 
to an assumption made in the experimental analysis on $|V_{us}f_1(0)|$, 
where the induced second-class form factor $g_2$ is set to be zero regardless of
broken $SU(3)$ symmetry.
We report on this matter and then estimate the possible value of $g_2(0)$, which is
evaluated from the experimental decay rate with our lattice result of $f_1(0)$ 
under the first-row CKM-unitarity condition.

\end{abstract}


\maketitle

\newpage

 
\section{Introduction}
\label{Sec:Sec1}
The Cabibbo-Kobayashi-Maskawa (CKM) matrix elements
are fundamental parameters of the Standard Model. 
So far, the most stringent test of the CKM unitarity 
is provided by the first-row relation $|V_{ud}|^2+|V_{us}|^2+|V_{ub}|^2=1-\Delta_{\rm CKM}$,
which can be examined accurately as $\Delta_{\rm CKM}=0.005(5)$~\cite{PDG}.
Since $|V_{ub}|^2\simeq 1\times 10^{-5}$ is negligibly small  in the first-row relation, 
the elements $|V_{ud}|$ and $|V_{us}|$ play crucial roles in this unitarity test. 
Combined with the experimental data on the semileptonic kaon ($K_{l3}$) decays,
the latest lattice calculations of the $K_{l3}$ form factor greatly contribute to 
the determination of $|V_{us}|$, which is one of the key elements~\cite{Aoki:2016frl}. 

The $\Delta S=1$ semileptonic hyperon decays offer an alternative way to 
extract $|V_{us}|$ accurately.  As we will explain later, however, the determination of $|V_{us}|$ from 
the semileptonic hyperon decays suffers from larger theoretical uncertainties than those of the $K_{l3}$ decay. 

The rate of $B_1 \rightarrow B_2 l\bar{\nu}$ semileptonic decay 
($B_1 \rightarrow B_2$ beta decay) is given by
%
%
\begin{multline}
\Gamma=\frac{G_F^2}{60\pi^3}(M_{B_1}-M_{B_2})^5(1-3\delta)|V_{us}|^2|f_1(0)|^2 \cr
{}\times(1+\Delta_{\rm RC})\left[
1+3\left|
\frac{g_1(0)}{f_1(0)}
\right|^2+\cdot\cdot\cdot
\right],
\label{Eq:DecayRate}
\end{multline}
where $M_{B_1}$ ($M_{B_2}$) denotes the rest mass
of the initial (final) state.
The Fermi constant $G_F$, which can be measured from 
the muon lifetime, already includes some electroweak radiative corrections~\cite{Cabibbo:2003cu}. 
The remaining radiative corrections to the decay rate are approximately 
represented by $\Delta_{\rm RC}$~\cite{Garcia:1985xz}. 
The ellipsis can be expressed in terms of a power series in the small parameter
$\delta=(M_{B_1}-M_{B_2})/(M_{B_1}+M_{B_2})$, which is regarded as a size of flavor $SU(3)$ 
breaking~\cite{Gaillard:1984ny}. 
The first linear term in $\delta$ is given by $-4\delta[g_2(0)g_1(0)/f_1(0)^2]$, 
where $f_1(0)$, $g_1(0)$, and $g_2(0)$ denote the vector, axial-vector and weak electricity
form factors at vanishing momentum transfer, respectively~\footnote{
Conventionally, $(M_{B_1}-M_{B_2})/M_{B_1}$ is adopted in Eq.(\ref{Eq:DecayRate})
to be the small parameter $\delta$~\cite{{Cabibbo:2003cu},{Gaillard:1984ny}}. 
However, our definition of the $SU(3)$-breaking parameter, 
$\delta=(M_{B_1}-M_{B_2})/(M_{B_1}+M_{B_2})$, is theoretically preferable
for considering the time-reversal symmetry on the matrix elements of the
hyperon beta decays in lattice QCD calculations~\cite{{Guadagnoli:2006gj},{Sasaki:2008ha}}.
Accordingly, a factor of $(M_{B_1}+M_{B_2})/M_{B_1}$ is different
in definitions of the $g_2$ form factor in comparison to those adopted in experiments.
}.
An essential difference from the case of the $K_{l3}$ decay is that the axial-vector transition, namely 
couplings $g_1(0)$ and $g_2(0)$, also contribute to the decay rate.

According to Weinberg's classification~\cite{Weinberg:1958ut}, the $g_2$ form factor is known as
one of the second-class form factors, which should be identically zero in the exact $SU(3)$ symmetry 
limit within the Standard Model~\cite{Cabibbo:2003cu}. 
Therefore, the nonzero value of $g_2(0)$ would be induced at first order in $SU(3)$ breaking.
It thus turns out that the term proportional to $\delta$ can be safely ignored 
as small as ${\cal O}(\delta^2)$~\cite{Gaillard:1984ny}. Recall that the expected
size of the second-order corrections is a few percent level since the mass
splittings among octet baryons is typically of the order of 10-15\%.
The absolute value of $g_1(0)/f_1(0)$ can be determined by measured 
asymmetries such as the electron-neutrino correlation~\cite{{Cabibbo:2003cu},{Gaillard:1984ny}}. 
Therefore, theoretical knowledge of $f_1(0)$, whose square is proportional to the decay rate, 
is crucial for obtaining $|V_{us}|$ from experimental measurements of the rate for the hyperon beta decays.

In the iso-spin limit ($m_u=m_d=m_{ud}$), all $\Delta S=1$ semileptonic hyperon 
decays can be classified in four types of beta decay: $\LtoN$, $\StoN$, $\XtoL$ and $\XtoS$
beta decays. Their values of $f_1(0)$ are known to be equal to the $SU(3)$ Clebsch-Gordan 
coefficients (denoted as $f_1^{\rm SU(3)}$ hereafter) in the exact $SU(3)$ 
symmetry limit ($m_{ud}=m_s$)~\cite{Cabibbo:2003cu}. 
However, in the real world, the $SU(3)$ symmetry is largely broken. 
Thus, a theoretical estimate of $SU(3)$ breaking-effects on the vector coupling $f_1(0)$ is primarily required 
for the precise determination of $|V_{us}|$ from the experimental rate of hyperon beta decays. 

Here, the hyperon vector coupling $f_1(0)$ can be parametrized using the value of $f_1^{\rm SU(3)}$ as below
%
%
\be
f_1(0)=f_1^{\rm SU(3)}(0)\left(1+\Delta f \right),
\ee
where $\Delta f$ represents full $SU(3)$-breaking corrections on $f_1(0)$.
According to the Ademollo-Gatto theorem (AGT)~\cite{Ademollo:1964sr}, 
$\Delta f$ starts only at the second order in the $SU(3)$ breaking. 
Therefore, $\Delta f$ is expected to be a few-percent correction at most.
However, either the size or the sign of $\Delta f$ is still controversial among various 
theoretical studies~\cite{Mateu:2005wi}. 

For two of the four independent semileptonic hyperon decays: $\Sigma^{-}\rightarrow nl^-\bar{\nu_l}$ 
(denoted as $\StoN$) and $\Xi^0\rightarrow\Sigma^{+}l^-\bar{\nu_l}$ (denoted as $\XtoS$), 
we reported the first results for the hyperon vector coupling $f_1(0)$ determined 
from fully dynamical lattice QCD with a range of pion masses down to $M_\pi \approx 330$ MeV 
at a single lattice spacing ($a\approx 0.114$ fm)~\cite{Sasaki:2012ne}. 
Our results show that the signs of $\Delta f$ are negative and its sizes are estimated 
as about 3\% for both $\StoN$ and $\XtoS$ beta decays. 
It is consistent with what was reported in earlier quenched 
lattice studies~\cite{{Guadagnoli:2006gj},{Sasaki:2008ha}}
and preliminary results from the mixed action calculation~\cite{Lin:2008rb}
and the dynamical improved Wilson fermion calculation~\cite{Gockeler:2011se}.
Although a recent unquenched lattice calculation~\cite{Shanahan:2015dka}
predicts more significant $SU(3)$ breaking-effects on $f_1(0)$ 
in all four channels, the signs of $\Delta f$ still agree with our results. 

In this paper, we extend our earlier work~\cite{Sasaki:2012ne} in order to examine 
possible systematic uncertainties including lattice artifacts due to the finite lattice spacing. 
We particularly determine the hyperon vector coupling $f_1(0)$ from fully dynamical lattice QCD 
with a range of pion masses down to $M_\pi \approx 290$ MeV
at a second value of the lattice spacing ($a\approx 0.086$ fm), which 
allows us to perform a continuum extrapolation. 

This paper is organized as follows: 
In Sec.\ref{Sec:Sec2}, we first summarize simulation parameters
in 2+1 flavor ensembles generated by the RBC and UKQCD Collaborations
with domain-wall fermions and the Iwasaki gauge action 
at two different lattice spacings, 
and then we describe the lattice method for calculating the target
form factor of the hyperon beta decay in order to determine the hyperon vector
coupling $f_1(0)$.
The numerical results are presented in Sec.~\ref{Sec:Sec3}.
We discuss in detail the $q^2$ interpolation of the form factor
and also the chiral-continuum extrapolation of the hyperon 
vector couplings for both $\StoN$ and $\XtoS$ beta decays.
Finally, we close with a brief summary and 
our conclusions in Sec.~\ref{Sec:Sec4}

\section{Simulation details}
\label{Sec:Sec2}

%
%
\begin{table}[ht]
\caption{
Details of the gauge ensembles: gauge coupling $\beta=6/g^2$, simulated masses
for the light ($am_{ud}$) and strange ($am_s$) quarks, the range, where measurements 
were made in this study, in molecular-dynamics (MD) time, 
the number of trajectory separation between each 
measured configuration ($N_{\rm sep}$),
the number of gauge configurations ($N_{\rm conf}$), and the number of different source positions
used on each configuration ($N_{\rm src}$), respectively.
The total number of measurements is therefore $N_{\rm conf}\times N_{\rm src}$.
For $\beta=2.13$, we include additional numerical simulations, which aim to more than 
double the total number of measurements in comparison to our earlier work~\cite{Sasaki:2012ne}.
}\label{Tab:Summary_DWFConf}
\begin{ruledtabular}
\begin{tabular}{cccrccc}
\hline
$\beta$  & $am_{ud}$  & $am_{s}$ & \multicolumn{1}{c}{MD range} & $N_{\rm sep}$ & 
\multicolumn{1}{c}{$N_{\rm conf}$} & $N_{\rm src}$\cr
\hline
2.13 & 
0.005 & 0.040 & 940-5720 & 20 & 240  & 8\cr
&0.010 & 0.040 & 5060-7440 & 20 & 120 & 8 \cr
&0.020 & 0.040 &  1890-3470 & 20 & 80 & 8\cr
\hline
2.25 & 
0.004 & 0.030 & 1000-3380 & 20 & 120 & 8\cr
& 0.006 & 0.030 & 1000-3380 & 20 & 120 & 8\cr
& 0.008 & 0.030 & 580-2960 & 20 & 120 & 8\cr
\hline
\end{tabular}
\end{ruledtabular}
\end{table}

In this paper, we use 2+1 flavor domain-wall fermions (DWF) lattice QCD ensembles generated
by the RBC and UKQCD Collaborations at two gauge couplings $\beta=2.13$~\cite{Allton:2008pn}
and $\beta=2.25$~\cite{Aoki:2010dy}. The former corresponds to a lattice spacing 
$a\approx0.114$ fm (coarse), while the latter corresponds to $a\approx0.086$ fm (fine).
Therefore, their lattice sizes, $L^3\times T=24^3\times64$ and $32^3\times 64$, correspond 
to almost the same physical volumes ($La\approx 2.7$ fm). 
Details of the gauge ensembles are given in Table~\ref{Tab:Summary_DWFConf}.
For more details on these ensembles see Refs.~\cite{{Allton:2008pn},{Aoki:2010dy}}.

The dynamical light and strange quarks are described by DWF actions with fifth-dimensional 
extent $L_5=16$ and the domain-wall height of $M_5=1.8$ for both ensemble sets. 
A brief summary of our simulation parameters with 2+1 flavor DWF 
ensembles appears in Table~\ref{Tab:Summary_DWFSim}.
Hereafter, the ensembles generated at $\beta=2.13$ are labeled as the $24^3$ lattice data, while
the ensembles generated with $\beta=2.25$ are labeled as the $32^3$ lattice data. 
Our previous results of $f_1(0)$ calculated from the $24^3$ ensembles with less number of measurements 
were published in Ref~\cite{Sasaki:2012ne}, while preliminary results of $f_1(0)$ obtained from 
the $32^3$ ensembles were first reported in Ref~\cite{Sasaki:2014osa}.

%
%
%
\begin{table*}[ht]
\caption{
Summary of simulation parameters in 2+1 flavor DWF ensembles with two different 
lattice spacings: gauge coupling $\beta=6/g^2$, lattice size, fifth-dimensional extent ($L_5$),
domain-wall height ($aM_5$), simulated masses for the light ($am_{ud}$)
and strange ($am_s$) quarks, the residual mass ($am_{\rm res}$), 
the physical strange quark mass ($am_s^{\rm phys}$) and inverse lattice spacing.
Each ensemble set of gauge configurations has been generated by the RBC and UKQCD
Collaborations; see Refs.~\cite{{Allton:2008pn},{Aoki:2010dy}} for further details.
}\label{Tab:Summary_DWFSim}
\begin{ruledtabular}
\begin{tabular}{lccccccllc}
\hline
 & $\beta$& $L^3\times T$  & $L_5$ & $aM_5$&$am_{ud}$  & $am_{s}$ 
& \multicolumn{1}{c}{$am_{\rm res}$} 
& \multicolumn{1}{c}{$am_s^{\rm phys}$} & $1/a$ [GeV] \cr
\hline
$24^3$ lattice &2.13 & $24^3\times 64$ & 16 & 1.8
&0.005, 0.010, 0,020 & 0.040 & 0.003 152(43) & 0.0348(11) & 1.73(3) \cr
$32^3$ lattice & 2.25 & $32^3\times 64$ & 16 & 1.8
& 0.004, 0.006, 0.008 & 0.030 & 0.000 666 4(76)
& 0.0273(7) & 2.28(3) \cr
\hline
\end{tabular}
\end{ruledtabular}
\end{table*}

\subsection{two-point correlation function}
\label{Sec:Sec2_twopt}

In order to compute baryon masses or beta-decay matrix elements, we
use the following spin-1/2 baryon interpolating operator:
%
%
\begin{multline}
(\eta_X^{S})_{ijk}(t, {\bm p})\\
{}=\sum_{{\bm x}}e^{-i{\bm p}\cdot{\bm x}}
\varepsilon_{abc}\left[
q_{a,i}^T({\bm y}_1,t)C\gamma_5q_{b,j}({\bm y}_2,t)\right]
q_{c,k}({\bm y}_3,t) \\
{}\times \phi({\bm y}_1-{\bm x})\phi({\bm y}_2-{\bm x})\phi({\bm y}_3-{\bm x}),
\end{multline}
where $C$ is the charge conjugation matrix defined as $C=\gamma_4\gamma_2$
and the index $X\in\{B_1, B_2\}$ distinguishes between the initial ($B_1$)
and final ($B_2$) states in the $B_1\rightarrow B_2$ beta decay.
The superscript $T$ denotes a transposition and the indices $abc$ and $ijk$
label color and flavor, respectively. 
The superscript $S$ of the interpolating operator $\eta$ specifies the smearing
for the quark propagators. In this study, we use two types of smearing function $\phi$:
the local function as $\phi({\bm x}_i-{\bm x})=\delta({\bm x_i}-{\bm x})$ and
the Gaussian-type distribution function. For the gauge invariance of the two-point function, 
${\bm x}_1={\bm x}_2={\bm x}_3={\bm x}_{\rm src}$ should be kept. 

We construct two types of the two-point function for octet baryon states
from the Gaussian-smeared quark fields at the source location
%
%
\be
C_X^{SG}(t-t_{\rm src},{\bm p})=\frac{1}{4}{\rm Tr}
\left\{
{\cal P}_{+}\langle 
\eta_X^{S}(t, {\bm p})
\bar{\eta}_X^{G}(t_{\rm src},-{\bm p})
\rangle
\right\},
\ee
where $S=L$ (local) or $G$ (Gaussian) stands for a type of smearing at the sink~\footnote{
As pointed out in Ref.~\cite{Sasaki:2007gw}, it is rather expensive to make the Gaussian smeared 
interpolating operator projected onto a specific finite momentum at the source location.
However, it is sufficient to project only the sink operator onto the desired momentum
by virtue of momentum conservation. Thus, the quark fields at the source location ${\bm x}_{\rm src}$
are not projected on to any specific momentum in this calculation.
}. 
A projection operator ${\cal P}_{+}=\frac{1+\gamma_4}{2}$ can eliminate
contributions from the opposite-parity state for $|{\bm p}|=0$~\cite{{Sasaki:2001nf},{Sasaki:2005ug}}. 
For the Gaussian smearing, we use gauge-covariant, approximately Gaussian-shaped 
smearing method~\cite{{Gusken:1989qx},{Alexandrou:1992ti}}, where
there are two parameters: the number of times the smearing kernel acts on
the quark fields ($N_G$) and the width of the Gaussian ($W_G$) that results in $N_G\rightarrow \infty$. 
Details of these definitions, see Ref~\cite{Berruto:2005hg}. Our choice of smearing parameters
$\{ N_G, W_G\}=\{100, 7\}$ follows an optimal set determined in the previous studies of
the nucleon structure on the same ensembles~\cite{{Yamazaki:2009zq},{Yamazaki:2008py},{Aoki:2010xg},{Syritsyn:2009mx}}.  In this study, for the finite three momentum $\bm p$, we use
the four lowest nonzero momenta: $\bm p=2\pi/L\times(1,0,0)$, $(1,1,0)$, $(1,1,1)$,
and $(2,0,0)$ in both $24^3$ ensembles and 
$32^3$ ensembles. 

We use the local interpolating operators, ${\bar u}(x)\gamma_5 d(x)$ for the pion,
${\bar u}(x)\gamma_5 s(x)$ for the kaon and also ${\bar s}(x)\gamma_5 s(x)$ for the
$\eta_{s}$ state. In Table~\ref{Tab:Summary_meson_DWF}, we summarize
the results of these meson masses together with the fit range $[t_{\rm min}/a:t_{\rm max}/a]$
used in the fits. All fitted values are obtained from the conventional cosh fit
for the LG-type two-point correlation functions. Our simulated values of the pion mass 
range from 330 MeV to 557 MeV for the $24^3$ ensembles
and from 290 MeV to 393 MeV for the $32^3$ ensembles. 

As for the octet baryons ($N$, $\Sigma$, $\Xi$, $\Lambda$), 
we adopt the conventional spin-1/2 baryon operators as below
%
%
\begin{align*}
&\eta_{N}(x)=\varepsilon_{abc}\left[d_a^T(x)C\gamma_5 u_b(x)\right]d_c(x) \;\;{\rm or}\;\;
[u \leftrightarrow d],\cr
&\eta_{\Sigma}(x)=\varepsilon_{abc}\left[u_a^T(x)C\gamma_5 s_b(x)\right]u_c(x)  \;\;{\rm or}\;\;
[u \leftrightarrow d],\cr
&\eta_{\Xi}(x)=\varepsilon_{abc}\left[s_a^T(x)C\gamma_5 u_b(x)\right]s_c(x)  \;\;{\rm or}\;\; 
[u \leftrightarrow d],\cr
&\eta_{\Lambda}(x)=\frac{\varepsilon_{abc}}{\sqrt{6}}\left\{\left[d_a^T(x)C\gamma_5 s_b(x)\right]u_c(x)
\right.\\
&+\left.\left[s_a^T(x)C\gamma_5 u_b(x)\right]d_c(x)
-2\left[u_a^T(x)C\gamma_5 d_b(x)\right]s_c(x)
\right\},
\end{align*}
where $[u \leftrightarrow d]$ indicates that other charge state's operators are obtained via 
the exchange $u \leftrightarrow d$.
For spin-1/2 baryon masses, 
all results obtained from the single exponential fit are tabulated in Table~\ref{Tab:Summary_baryon_DWF}.
The errors quoted in both Tables~\ref{Tab:Summary_meson_DWF}
and \ref{Tab:Summary_baryon_DWF} represent only the statistical errors given by the jackknife analysis.
In later analysis of form factors, both LG and GG-type two-point correlation functions are used. 
Therefore, the fitted masses obtained from the GG-type correlators
are also included in Table~\ref{Tab:Summary_baryon_DWF}. Both results are consistent with
each other within their statistical errors, while the fit range for the GG-type correlators starts slightly closer to 
the source. In the later discussion, we use the baryon masses obtained from the LG-type correlators.

%
%
  \begin{figure}[hb]
  \centering
  \includegraphics*[width=.48\textwidth]{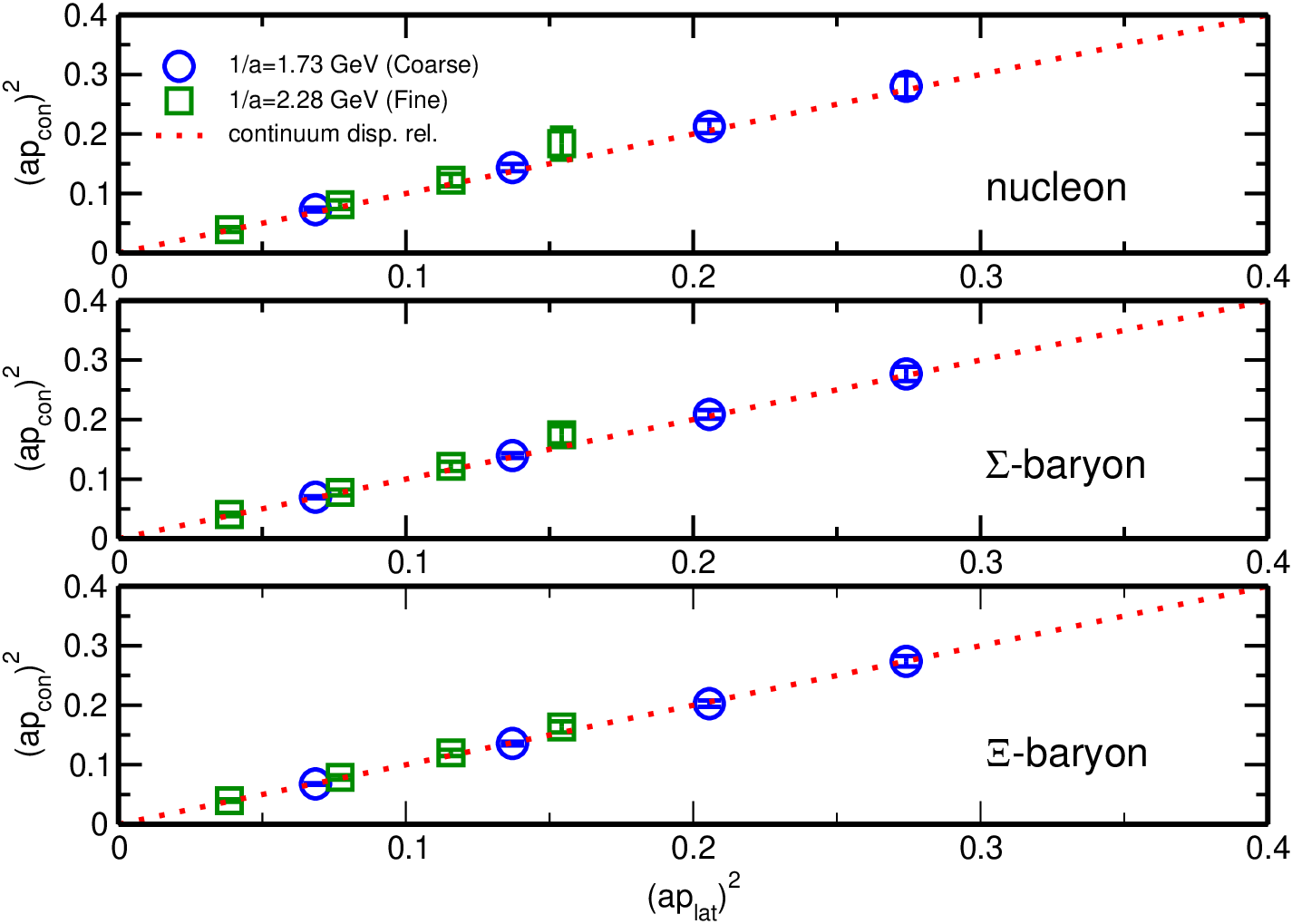} 
  \caption{
  Check of the dispersion relation for the nucleon (upper panel), $\Sigma$-baryon (middle panel)
  and $\Xi$-baryon (lower panel). Open circle (diamond) symbols are results from the $24^3$ ($32^3$) 
  ensembles with the lightest quark mass of $am_{ud}=0.005$ (0.004). The variables $p_{\rm con}^2$ 
  and $p^2_{\rm lat}$ appearing on the x-axis and y-axis are defined in text.
  For comparison, the continuum dispersion relation is denoted as the dotted line in each panel.
  }
  \label{fig:Disp}
  \end{figure}
%

%
%
\begin{table*}[ht]
\caption{
Mass spectrum of the pion, kaon and $\eta_{s}$-meson in lattice units.
All meson masses are computed by using the LG-type correlation functions.
}\label{Tab:Summary_meson_DWF}
\begin{ruledtabular}
\begin{tabular}{cclclclccc}
\hline
$\beta$ & $am_{ud}$
& \multicolumn{1}{c}{$aM_\pi$} & Fit range  
& \multicolumn{1}{c}{$aM_K$} & Fit range
& \multicolumn{1}{c}{$aM_{\eta_{s}}$} & Fit range & $N_{\rm meas}$ & Type 
\cr
\hline
2.13 & 0.005 & 0.1908(7) & [27:39] & 0.3327(6) & [27:39] & 0.4318(4) & [27:39]
& 240$\times$8 & LG
\cr
& 0.010 & 0.2436(9) & [26:40] & 0.3514(8) &  [26:40] & 0.4351(6) &  [26:40] 
& 120$\times$8 & LG
\cr
& 0.020 & 0.3219(11) & [26:40] & 0.3839(10) & [26:40] & 0.4380(9) & [26:40]
& 80$\times$8 & LG
\cr
\hline
2.25 
&0.004 & 0.1273(6) & [20:46] & 0.2436(7) & [20:46] & 0.3213(4) & [20:46]
& 120$\times$8& LG
\cr
&0.006  & 0.1511(5) & [20:46] & 0.2500(5) & [20:46] & 0.3214(4) & [20:46]
& 120$\times$8& LG
\cr
&0.008  & 0.1722(6) & [22:44] & 0.2578(5) & [22:44] & 0.3227(5) & [22:44]
& 120$\times$8& LG
\cr
\hline
\end{tabular}
\end{ruledtabular}
\end{table*}
%

%
%
\begin{table*}[ht]
\caption{
Mass spectrum of the nucleon, $\Sigma$, $\Xi$, and $\Lambda$-baryon in lattice units.
All baryon masses are computed by using both LG-type and GG-type correlation functions.
}\label{Tab:Summary_baryon_DWF}
\begin{ruledtabular}
\begin{tabular}{cclclclclccc}
\hline
$\beta$ & $am_{ud}$
& \multicolumn{1}{c}{$aM_{N}$} & Fit range 
& \multicolumn{1}{c}{$aM_{\Sigma}$} & Fit range
& \multicolumn{1}{c}{$aM_{\Xi}$}  & Fit range 
& \multicolumn{1}{c}{$aM_{\Lambda}$} & Fit range
& $N_{\rm meas}$ & Type
\cr
\hline
2.13 
&0.005 & 0.660(4) & [7:13] & 0.773(3) & [7:13] & 0.829(2) & [7:13] & 0.738(3) & [7:13] 
& 240$\times$8 & LG
\cr
&0.010 & 0.725(4) & [6:18] & 0.809(3) & [6:18] & 0.856(2) & [6:18] & 0.784(3) & [6:18] 
& 120$\times$8 & LG
\cr
&0.020 & 0.813(5) & [7:17] & 0.864(4) & [7:17] & 0.892(3) & [7:17] & 0.848(4) & [7:17] 
& 80$\times$8 & LG
\cr
\hline
2.13
&0.005 &  0.650(5)  & [6:13] &  0.763(4) &[6:13] &  0.822(3)  & [6:13]&  0.730(3)  & [6:13] 
& 240$\times$8 & GG \cr
&0.010 & 0.719(5)  & [5:14] & 0.805(4) &[5:14] &  0.854(3)  & [6:14]&  0.780(4)  & [5:14] 
& 120$\times$8 & GG \cr
&0.020 & 0.807(5)  & [5:11] &  0.859(5) &[5:11] &  0.890(4) & [5:11]& 0.845(5) & [5:11] 
& 80$\times$8 & GG \cr
\hline\hline
2.25 
&0.004  & 0.491(6) &[10:28]&  0.579(4) &[10:28]&  0.620(3) &[10:28]&  0.551(3) &[10:28] 
& 120$\times$8 & LG
\cr
&0.006  & 0.501(5) &[10:28]&  0.581(4) &[10:28]&  0.624(3) &[11:28]&  0.558(4) &[11:28] 
& 120$\times$8 & LG
\cr
&0.008 & 0.524(4) & [11:24]& 0.594(4) & [11:24]& 0.633(2) &[11:24] & 0.574(3) &[11:24] 
& 120$\times$8 & LG
\cr
\hline
2.25 
&0.004  & 0.490(5) &[8:25]&  0.576(4) &[8:25]&  0.617(3) &[8:25]&  0.548(4) &[8:25] 
& 120$\times$8 & GG
\cr
&0.006  & 0.501(5) &[8:25]&  0.579(4) &[8:25]&  0.620(3) &[8:25]&  0.555(3) &[8:25] 
& 120$\times$8 & GG
\cr
&0.008 & 0.518(4) & [9:25]& 0.595(4) & [8:25]& 0.631(3) &[8:25] & 0.568(4) &[9:25] 
& 120$\times$8 & GG
\cr
\hline
\end{tabular}
\end{ruledtabular}
\end{table*}

We also measure the baryon energies $E_{X}({\bm p})$ ($X=N, \Sigma, \Xi$)
from the LG-type correlators with four nonzero momenta $|\bm p|\neq 0$.
As shown in Fig.~\ref{fig:Disp}, the measured energies $E_{X}({\bm p})$ are 
well satisfied with the continuum dispersion relation on both $24^3$ and $32^3$ ensembles. 
The vertical axis shows the momentum squared defined through
the relativistic continuum dispersion relation as $p_{\rm con}^2=E_X^2-M_X^2$ for
$X=N, \Sigma$ and $\Xi$, while the horizontal axis is the momentum squared defined by
the lattice momentum $p^2_{\rm lat}=(2\pi/aL)^2 \times n$ ($n=1, 2, 3, 4$).
As typical examples, we plot the results for the nucleon (upper panel), $\Sigma$-baryon (middle panel) 
and $\Xi$-baryon (lower panel), that are calculated with the $24^3$ ($32^3$) ensembles 
at the lightest quark mass of $am_{ud}=0.005$ (0.004).

The evaluation of momentum transfer $q^2$ for the 
$\Sigma\rightarrow N$ and $\Xi \rightarrow \Sigma$ beta decays requires 
precise knowledge of the baryon energies $E_{X}({\bm p})$ in later analysis.
However, in general, the two-point correlation functions have higher statistical noise 
for the larger momentum. Instead of actually measured values, 
we thus use an estimation of the baryon energies $E_{X}({\bm p})$ 
through the continuum dispersion relation with the rest masses $M_X$, 
that are most precisely determined, in our whole analysis.

\subsection{Three-point correlation functions}
\label{Sec:Sec2_threept}

The general form of the weak matrix element for semileptonic hyperon decay  
$B_1 \rightarrow B_2 l\bar{\nu}$ is composed of the vector and axial-vector
transitions, $\langle B_2(p^\prime)|V_{\alpha}(x)+A_{\alpha}(x)|B_1(p)\rangle$,
which are described by six form factors: the vector ($f_1$), weak-magnetism ($f_2$), and
induced scalar ($f_3$) form factors for the vector current, 
and the axial-vector ($g_1$), weak electricity ($g_2$), and induced pseudo-scalar
($g_3$) form factors for the axial current~\cite{Cabibbo:2003cu}. 

In this paper, we focus on the vector part of the weak matrix element:
%
%
\be
\langle B_2(p^\prime)|V_{\alpha}(x)|B_1(p)\rangle 
=\bar{u}_{B_2}(p^\prime){\cal O}_{\alpha}^V(q)u_{B_1}(p)e^{iq\cdot x}
\ee
with
\begin{multline}
{\cal O}_{\alpha}^V(q)= 
\gamma_\alpha f_1^{\Btob}(q^2) \cr
+ \sigma_{\alpha\beta}q_\beta\frac{f_2^{\Btob}(q^2)}{M_{B_1}+M_{B_2}}
+iq_\alpha \frac{f_3^{\Btob}(q^2)}{M_{B_1}+M_{B_2}},
\label{Eq:WME}
\end{multline}
where $q\equiv p-p^\prime$ is the momentum transfer between the initial state ($B_1$)
and the final state ($B_2$) which belong to the lightest $J^P=1/2^{+}$ $SU(3)$ octet
of baryons ($N, \Lambda, \Sigma, \Xi$). Recall that Eq.~(\ref{Eq:WME}) is given in the Euclidean 
metric convention (see Ref.~\cite{Sasaki:2008ha} for details). 

In order to calculate the weak matrix element on the lattice, 
we next define the finite-momentum three-point functions for the hyperon 
beta-decay process $B_1({\bm p})\rightarrow B_2({\bm p}^\prime)$:
%
%
\begin{multline}
C_\alpha^{B_{1}\rightarrow B_{2}}(t, {\bm p}^\prime, {\bm p})\\
{}=\frac{1}{4}{\rm Tr}
\left\{
{\cal P}_{+}\langle \eta_{B_2}(t_{\rm sink}, {\bm p}^\prime)
V_\alpha(t, {\bm q})\bar{\eta}_{B_1}(t_{\rm src}, -{\bm p})\rangle
\right\},
\end{multline}
where $V_\alpha$ denotes the local vector current, which is 
defined by $V_\alpha(x)=\bar{u}(x)\gamma_\alpha s(x)$ for $\Delta S=1$ decays.

We then calculate the following ratio constructed from the three-point function 
$C_\alpha^{B_{1}\rightarrow B_{2}}$ with two-point functions of $B_1$ and $B_2$ states:
%
%
\begin{widetext}
\be
{\cal R}_\alpha^{B_{1}\rightarrow B_{2}}(t, {\bm p}^\prime, {\bm p})=
\frac{C_\alpha^{B_{1}\rightarrow B_{2}}(t, {\bm p}^\prime, {\bm p})}
{C^{\rm GG}_{B_2}(t_{\rm sink}-t_{\rm src}, {\bm p}^\prime)}
\left[
\frac{C^{\rm LG}_{B_1}(t_{\rm sink}-t, {\bm p})C^{\rm GG}_{B_2}(t-t_{\rm src}, {\bm p}^\prime)
C^{\rm LG}_{B_2}(t_{\rm sink}-t_{\rm src}, {\bm p}^\prime)}
{C^{\rm LG}_{B_2}(t_{\rm sink}-t, {\bm p}^\prime)C^{\rm GG}_{B_1}(t-t_{\rm src}, {\bm p})
C^{\rm LG}_{B_1}(t_{\rm sink}-t_{\rm src}, {\bm p})}
\right]^{1/2},
\ee
\end{widetext}
which is a function of the current operator insertion time $t$ at the given values of 
momenta ${\bm p}^\prime$ and ${\bm p}$ for the initial and final states. 

In this study, we consider the hyperon beta-decay process $B_1({\bm p})\rightarrow B_2({\bm 0})$
at the rest flame of the final ($B_2$) state (${\bm p}^\prime={\bm 0}$), which
leads to ${\bm q}={\bm p}$. Therefore, the squared four-momentum transfer is given
by $q^2=2M_{B_2}(E_{B_1}({\bm p})-M_{B_1})-(M_{B_1}-M_{B_2})^2$. The energies
of the initial baryon states are simply abbreviated as $E_{B_1}$, hereafter.
In these kinematics, ${\cal R}_\alpha^{B_{1}\rightarrow B_{2}}(t, {\bm p}^\prime, {\bm p})$
is represented by a simple notation ${\cal R}_\alpha^{B_{1}\rightarrow B_{2}}(t, {\bm q})$,
which gives the following asymptotic values~\cite{Sasaki:2008ha} 
%
%
\begin{widetext}
\bea
{\cal R}_4^{B_{1}\rightarrow B_{2}}(t, {\bm q})&\rightarrow& \sqrt{\frac{E_{B_1}+M_{B_1}}{2E_{B_1}}} 
\left[f_1^{B_{1}\rightarrow B_{2}}(q^2)
-\frac{E_{B_1}-M_{B_1}}{M_{B_1}+M_{B_2}}f_2^{B_{1}\rightarrow B_{2}}(q^2)
-\frac{E_{B_1}-M_{B_2}}{M_{B_1}+M_{B_2}}f_3^{B_{1}\rightarrow B_{2}}(q^2)
\right],
\label{Eq:R_time}\\
{\cal R}_{i}^{B_{1}\rightarrow B_{2}}(t, {\bm q}) &\rightarrow& \frac{-iq_{i}}{\sqrt{2E_{B_1}(E_{B_1}+M_{B_1})}}
\left[f_1^{B_{1}\rightarrow B_{2}}(q^2)
-\frac{E_{B_1}-M_{B_2}}{M_{B_1}+M_{B_2}}f_2^{B_{1}\rightarrow B_{2}}(q^2)
-\frac{E_{B_1}+M_{B_1}}{M_{B_1}+M_{B_2}}f_3^{B_{1}\rightarrow B_{2}}(q^2)
\right] 
\label{Eq:R_space}
\eea
\end{widetext}
in the limit when the Euclidean time separation between all operators is large, $t_{\rm sink} \gg t \gg t_{\rm src}$
with fixed $t_{\rm src}$ and $t_{\rm sink}$.
Let us define the dimensionless ratios~\cite{Sasaki:2008ha}:
%
%
\bea
\Lambda_4^{B_{1}\rightarrow B_{2}}(t, {\bm q})&=&\sqrt{\frac{2E_{B_1}}{E_{B_1}+M_{B_1}
}}{\rm Re}\{{\cal R}_4^{B_{1}\rightarrow B_{2}}\}(t, {\bm q}), \nonumber\\
\label{Eq:Lambda_time} \\
\Lambda_S^{B_{1}\rightarrow B_{2}}(t, {\bm q})&=&\frac{\sqrt{2E_{B_1}(E_{B_1}+M_{B_1})}}{3},
\nonumber \\
&&\times \sum_{i=1,2,3}
\frac{{\rm Im}\{{\cal R}_i^{B_{1}\rightarrow B_{2}}(t, {\bm q})\}}{q_i},
\label{Eq:Lambda_space}
\eea
which are related to brackets that appear in Eqs.~(\ref{Eq:R_time}) and (\ref{Eq:R_space}).

For convenience in numerical calculations, instead of the vector form factor $f_1(q^2)$, 
we consider the so-called scalar form factor for the $B_1 \rightarrow B_2$ beta decay~\footnote{Here, we note 
that $f_1^{B_1\rightarrow B_2}=f_1^{B_2\rightarrow B_1}$, 
while $f_3^{B_1\rightarrow B_2}=-f_3^{B_2\rightarrow B_1}$}
%
%
\be
f_S^{B_1\rightarrow B_2}(q^2)=f_1^{B_1\rightarrow B_2}(q^2)+\frac{q^2}{M_{B_1}^2-M_{B_2}^2}
f_3^{B_1\rightarrow B_2}(q^2),
\ee
which become equal to the vector form factor $f_1^{B_1\rightarrow B_2}(q^2)$ in the exact $SU(3)$
limit ($m_{ud}=m_{s}$), where the second form factor $f_3(q^2)$ are prohibited from having
nonzero values because of the extended $G$ parity conservation regarding 
the $V$-spin symmetry~\cite{Weinberg:1958ut}.
Recall that the scalar form factor at $q^2=0$, $f_S(0)$, is identical to the vector coupling $f_1(0)$ 
even with the $SU(3)$ breaking. 

Finally, the scalar form factor $f_S(q^2)$ can be calculated from 
the following linear combinations of $\Lambda_4^{B_{1}\rightarrow B_{2}}(t, {\bm q})$ 
and $\Lambda_S^{B_{1}\rightarrow B_{2}}(t, {\bm q})$ as a plateau behavior,
%
%
\begin{multline}
\left(\frac{E_{B_1}-M_{B_2}}{M_{B_1}-M_{B_2}}\right)\Lambda^{B_{1}\rightarrow B_{2}}_{4}(t, {\bm q})\cr
-\left(\frac{E_{B_1}-M_{B_1}}{M_{B_1}-M_{B_2}}\right)\Lambda^{B_{1}\rightarrow B_{2}}_{S}(t, {\bm q})
\cr
= f_S^{B_{1}\rightarrow B_{2}}(q^2) + \cdot\cdot\cdot ,
\label{Eq:f_S}
\end{multline}
where the ellipses denote excited-state contributions that decay exponentially with the source-sink separation. 

%
%
\begin{table*}[ht]
\caption{
Results for the renormalized value of $|f_S(q_{\rm max}^2)|$, where $q_{\rm max}^2=-(M_{B_1}-M_{B_2})^2$
with $(B_1,B_2)=(\Xi,\Sigma)$ and $(\Sigma, N)$. The values of $q_{\rm max}^2$
are listed in units of ${\rm GeV}^2$. 
A common fit range for both $\NtoS$ and $\XtoS$ decays 
is taken as $[t_{\rm min}/a:t_{\rm max}/a]$=[4:8] ([5:10]) for $\beta=2.13$ (2.25).
}
\label{Tab:f0atqmax}
\begin{ruledtabular}
\begin{tabular}{cl l l l l }
\hline
 & &\multicolumn{2}{c}{$\StoN$} & \multicolumn{2}{c}{$\XtoS$} \cr
$\beta$ &$am_{ud}$ 
& \multicolumn{1}{c}{$q_{\rm max}^2$}
& \multicolumn{1}{c}{$|f^{\rm ren}_S(q_{\rm max}^2)|$} 
& \multicolumn{1}{c}{$q_{\rm max}^2$}
& \multicolumn{1}{c}{$|f^{\rm ren}_S(q_{\rm max}^2)|$}\cr
\hline
2.13 & 0.005 
          &$-0.0378(13)$   &   1.0205(62)
          &$-0.0097(5)$   &  0.9835(53) \cr
& 0.010   
       & $-0.0213(10)$ & $1.0083(25)$ 
	& $-0.0067(5)$ &  $0.9867(30)$ \cr
& 0.020   
	& $-0.0079(4)$ & $1.0024(5)$ 
	& $-0.0023(3)$ & $0.9898(12)$ \cr
\hline
2.25 & 0.004 
       &$-0.0405(40)$  &  $1.0356(77)$
	&$-0.0087(13)$ & $0.9845(61)$
       \cr
& 0.006   
       & $-0.0331(28)$ &  $1.0128(33)$
	& $-0.0097(14)$ &  $0.9859(28)$
	\cr
& 0.008   
	& $-0.0259(18)$ & $1.0087(20)$ 
	& $-0.0079(10)$  & $0.9895(22)$
	\cr
\hline
\end{tabular}
\end{ruledtabular}
\end{table*}
%

\section{Numerical Results}
\label{Sec:Sec3}

In this study, all three-point functions are calculated by the sequential source method
with a fixed source location~\cite{Sasaki:2003jh}. 
To increase statistics, we use four different time-slices ($t_{\rm src}$)
with two different spatial centers of the Gaussian smeared sources (${\bm x}_{\rm src}$).
Therefore, the total number of measurements on each configuration is eight. 
In the analysis, all 8 sets of three-point correlation functions and baryon two-point functions
are folded together to create the single-correlation functions, respectively. 
It can reduce possible autocorrelation among measurements. The source location is chosen 
at time slices of $t_{\rm src}=nT/4$ ($n=0,1,2,3$) with two 
spatial centers of the Gaussian smeared source at 
${\bm x}_{\rm src}=(\frac{mL}{4},\frac{mL}{4},\frac{mL}{4})$ where $m=n$ or $m=3-n$.  
We use the source-sink separation of 12(15) in lattice units for the $24^3$ ($32^3$) ensembles, 
which is large enough to suppress the excited state contributions~\cite{{Yamazaki:2009zq},{Syritsyn:2009mx}}.

\subsection{Scalar form factor $f_S(q^2)$ at $q^2=q^2_{\rm max}$}
\label{Sec:Sec3_f0max}

In the vector matrix element, only the time component of the vector current, namely 
the three-point correlation function $C_4^{\Btob}(t, {\bm q})$, is prevented from vanishing 
at zero three-momentum transfer $|{\bm q}|=0$, by the kinematics~\cite{Sasaki:2003jh}.
Thus, for the case of ${\bm q}={\bm 0}$, Eq.~(\ref{Eq:f_S}) reduces to a simple relation with
the scalar form factor at specific four-momentum transfer as
%
%
\be
\Lambda_4^{B_{1}\rightarrow B_{2}}(t,{\bm 0})=
f_S^{B_{1}\rightarrow B_{2}}(q_{\rm max}^2)+\cdot\cdot\cdot,
\ee
where $q_{\rm max}^2=-(M_{B_1}^2-M_{B_2}^2)$. Recall that the lattice operators receive finite renormalizations relative to their continuum counterparts in general. The local vector current $\bar{q}_f(x)\gamma_\alpha q_{f^\prime}(x)$ ($f, f^\prime$ denote flavor indices), that
is not the conserved one on the lattice, needs the vector renormalization factor $Z_V^{\bar{f}f^\prime}$. 
Thus, the renormalized value of the form factors ($k=1,2,3$ and $S$)
%
%
\be
f_k^{{\rm ren}}(q^2)= Z_V^{\bar{u}s}f_k(q^2)
\ee
requires some independent estimation of $Z_V^{\bar{u}s}$. Here,
we may calculate $Z_V^{\bar{u}s}=Z_V^{\bar{s}u}$  through the following relation:
%
%
\be
Z_V^{\bar{u}s}=Z_V^{\bar{s}u}=\sqrt{Z_V^{\bar{u}{u}}Z_V^{\bar{s}{s}}},
\ee
where $Z_V^{\bar{u}{u}}$ and $Z_V^{\bar{s}{s}}$ can be obtained
with the help of the conserved current vector relation under the exact iso-spin symmetry.

In this context, the renormalized value of $|f_S(q^2)|$ at $q_{\rm max}^2=-(M_{B_1}-M_{B_2})^2<0$
can be precisely evaluated by the double ratio method 
proposed in Refs~\cite{{Guadagnoli:2006gj},{Hashimoto:1999yp}}, where
all relevant three-point functions are determined at zero three-momentum transfer ${\bm q}^2=0$.
The double ration is defined by 
%
%
\begin{widetext}
\be
{\cal R}_{W}(t)=\sqrt{\frac{C_4^{B_{1}\rightarrow B_{2}}(t, {\bm 0})
C_4^{B_{2}\rightarrow B_{1}}(t, {\bm 0})}
{C_4^{B_{1}\rightarrow B_{1}}(t, {\bm 0})
C_4^{B_{2}\rightarrow B_{2}}(t, {\bm 0})}}
\phantom{.}_{\overrightarrow{\mbox{\small $t_{\rm sink}\gg t \gg t_{\rm src}$}}}
\sqrt{Z_V^{\bar{u}u}Z_V^{\bar{s}s}}
 \left| f_S^{{B_1\rightarrow B_2}}(q_{\rm max}^2)\right|
 = \left| f_S^{{\rm ren}, {B_1\rightarrow B_2}}(q_{\rm max}^2)\right|,
\label{Eq:DoubleRatio}
\ee
\end{widetext}
where the three-point functions of $B_{1}\rightarrow B_{1}$ ($B_{2}\rightarrow B_{2}$) 
in the denominator of the double ratio are defined with the vector current
$V_4(x)=\bar{s}(x)\gamma_4 s(x)$ for $B_{1,2}=\Sigma^\pm$ and 
$V_4(x)=\bar{u}(x)\gamma_4 u(x)$ for $B_{1,2}=n, \Xi^0$. 
The sign of $f_S(q_{\rm max}^2)$ can be read off from the sign of $\Lambda_4^{B_{1}\rightarrow B_{2}}(t,{\bm 0})$. The double ratio gives an asymptotic plateau corresponding to 
the renormalized value of $|f_S(q_{\rm max}^2)|$ in the middle region,  
between the source and sink points when the condition $t_{\rm sink}\gg t \gg t_{\rm src}$
is satisfied. 

%
%
  \begin{figure*}[ht]
  \centering
  \includegraphics*[width=.48\textwidth]{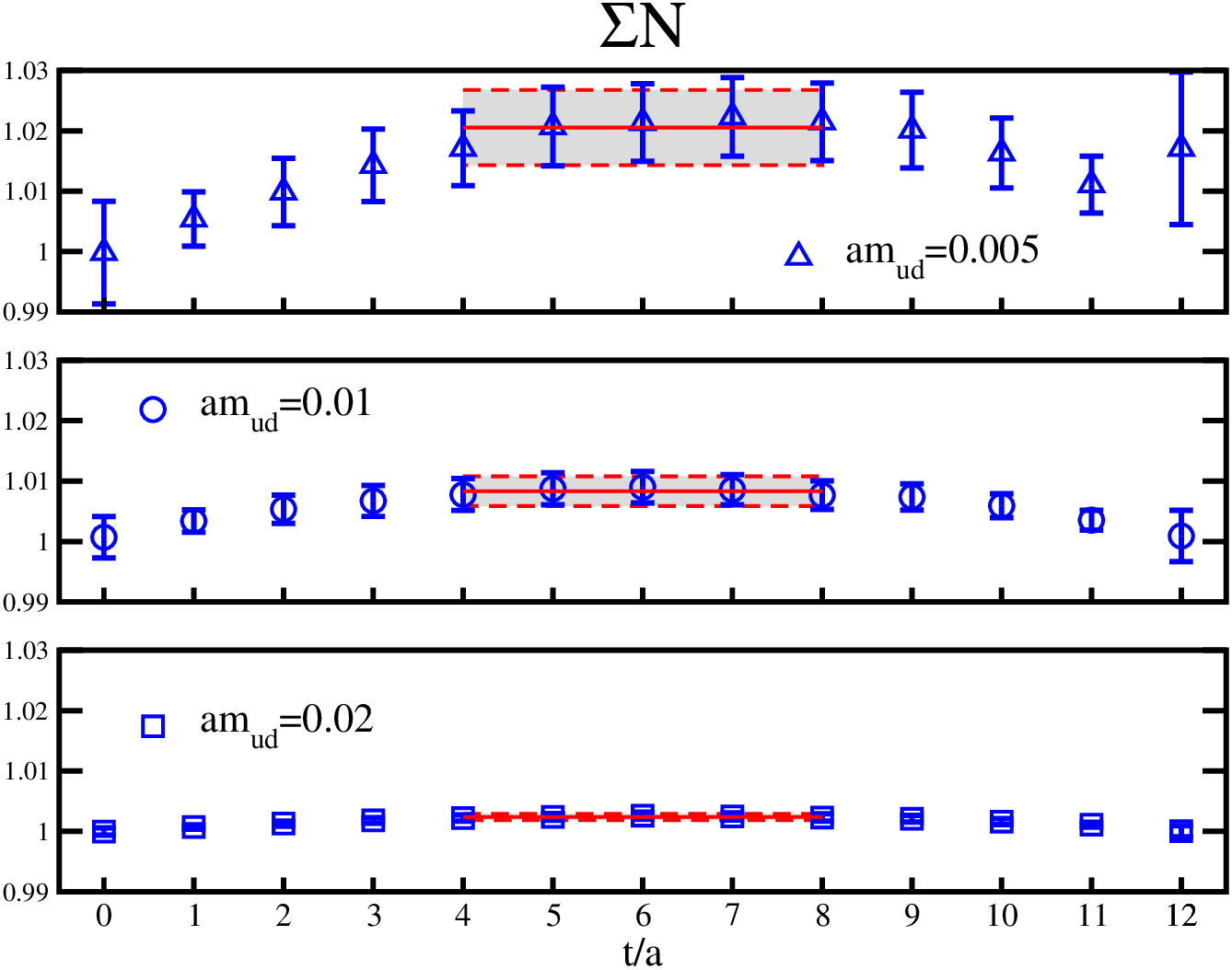} 
  \includegraphics*[width=.48\textwidth]{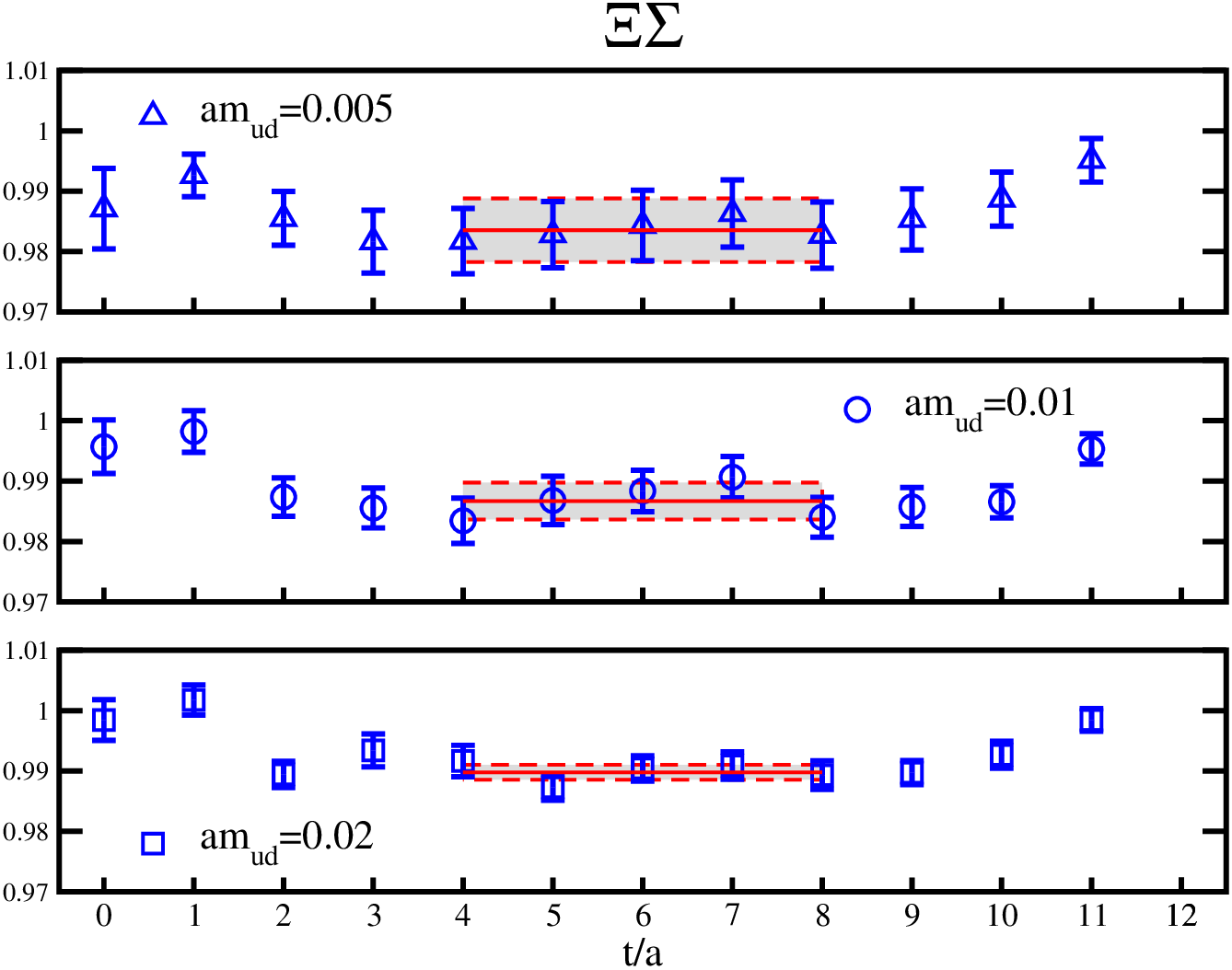} 
  \caption{The absolute value of $f^{\rm ren}_S(q_{\rm max}^2)$ computed on the $24^3$
  ensembles ($\beta=2.13$) as a function of the current insertion time-slice.
  The left (right) panel is for the $\StoN$ ($\XtoS$) beta decay. 
  In each plot, results for $am_{ud}=0.005$, 0.010, and 0.020 are plotted from top to
  bottom. The lines represent the average value (solid lines) and their 1 standard deviations (dashed
  lines) over range of $4\le t/a\le 8$.}
  \label{fig:f0max_b213}
  \end{figure*}
%

%
%
  \begin{figure*}[ht]
  \centering
  \includegraphics*[width=.48\textwidth]{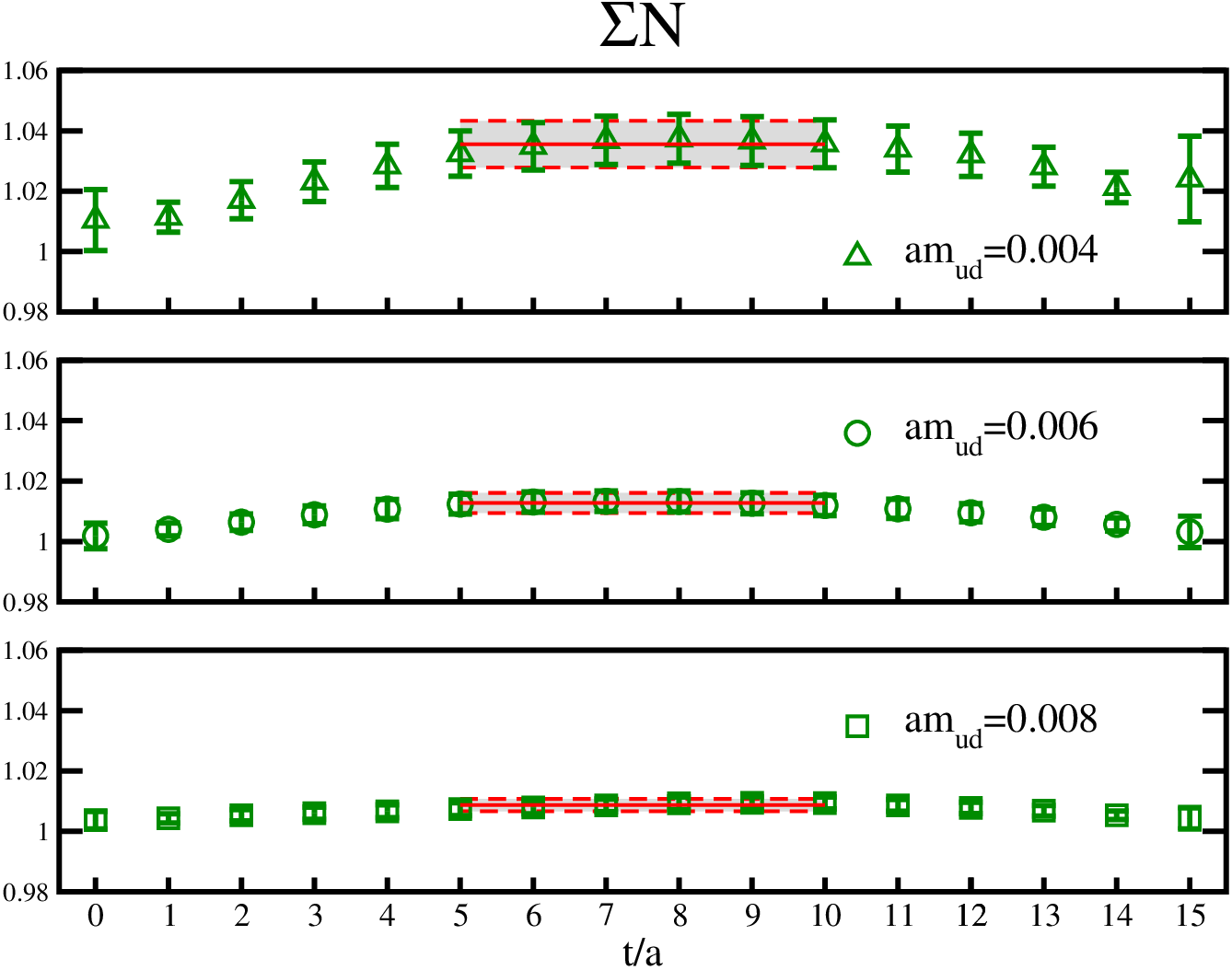} 
  \includegraphics*[width=.48\textwidth]{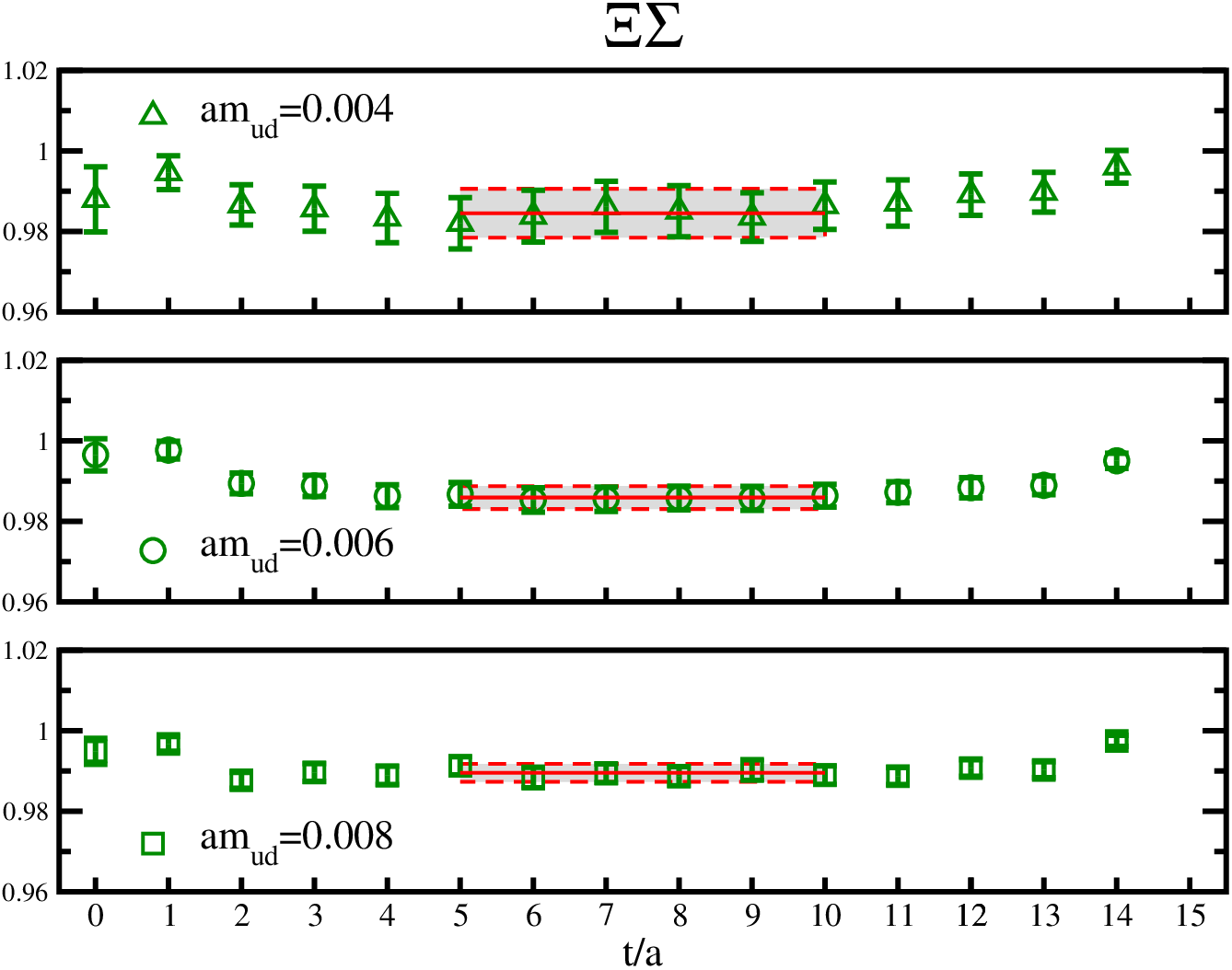} 
  \caption{The absolute value of $f^{\rm ren}_S(q_{\rm max}^2)$ computed on the $32^3$
  ensembles ($\beta=2.25$) as a function of the current insertion time-slice.
  The left (right) panel is for the $\StoN$ ($\XtoS$) beta decay. 
  In each panel, results for $am_{ud}=0.004$, 0.006, and 0.008 are plotted from top to
  bottom. The lines represent the average value (solid lines) and their 1 standard deviations (dashed
  lines) over range of $5\le t/a\le 10$.}
  \label{fig:f0max_b225}
  \end{figure*}
%

%
%
\begin{figure*}[ht]
  \centering
  \includegraphics*[width=.48\textwidth]{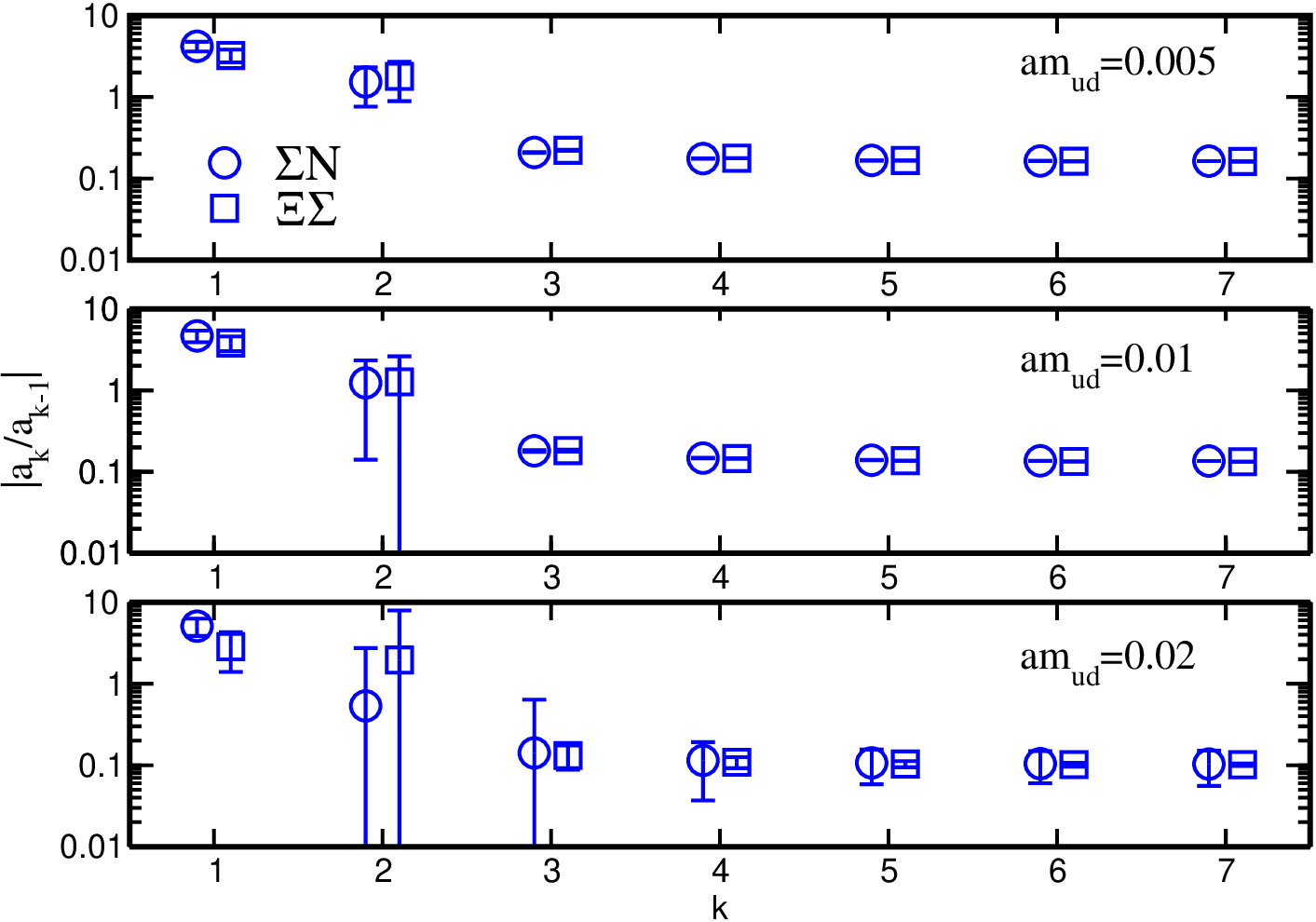} 
  \includegraphics*[width=.48\textwidth]{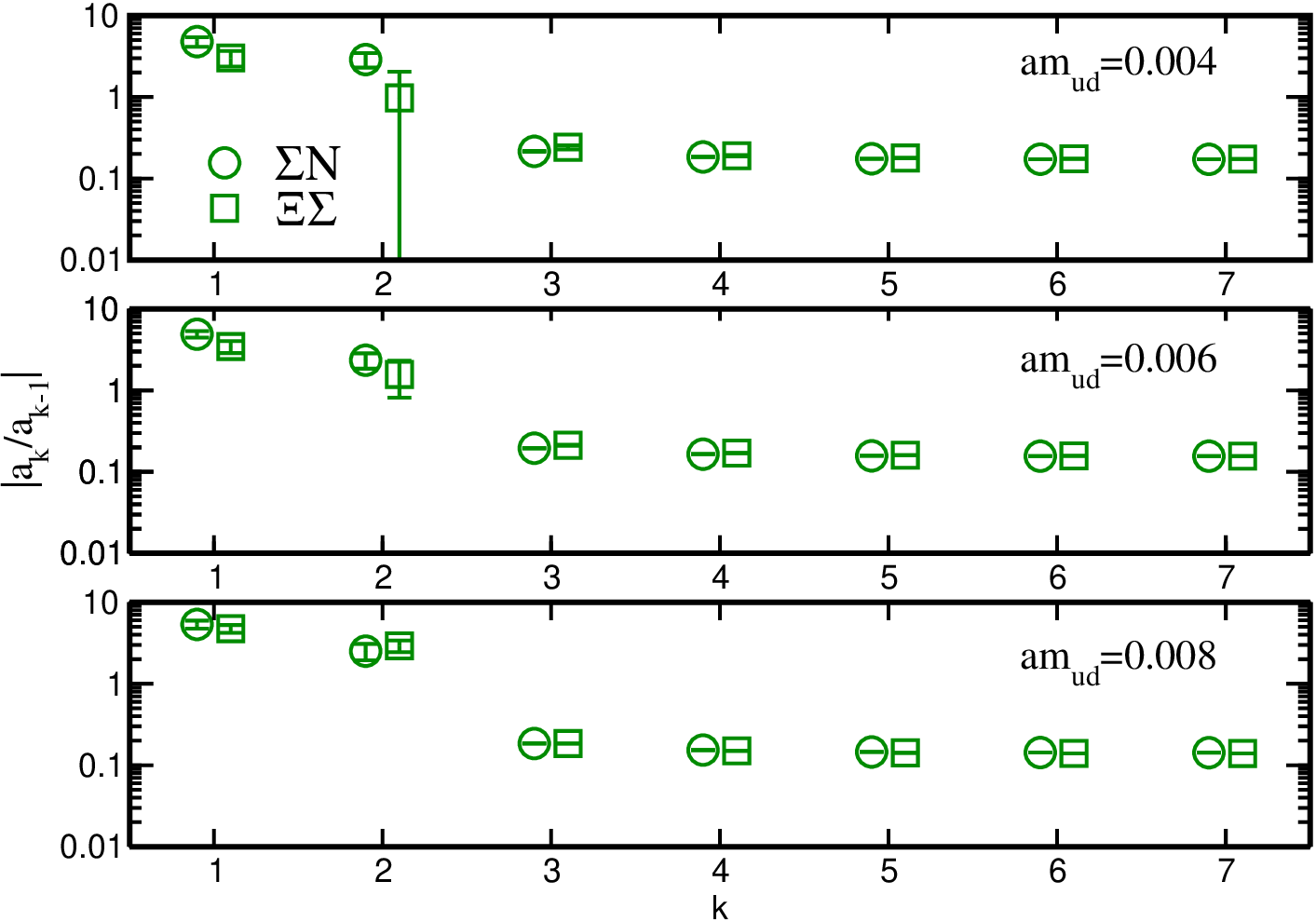} 
  \caption{Convergence behavior of the z-Exp fits for the $24^3$ (left panels) and $32^3$ (right
  panels) ensembles. The ratios of $|a_{k}/a_{k-1}|$ that
  are determined by fitting all nine data points with $k_{\rm max}=7$ reach a convergence value less
  than unity at $k\approx 3$. Open circles (squares) represent results for the $\StoN$ ($\XtoS$) beta decay.
  }
  \label{fig:svd_coeffs}
\end{figure*}

In Figs.~\ref{fig:f0max_b213} and \ref{fig:f0max_b225}, we plot the absolute value of the
renormalized $f_S(q_{\rm max}^2)$ as a function of the current insertion time-slice
for both $\StoN$ (left panels) and $\XtoS$ (right panels) beta decays. 
Good plateaus are observed in the middle region between the source and sink
points. In each plot, the lines represent the average value (solid lines) and their 1 standard deviations
(dashed lines) over range of $4\le t/a \le 8$ ($5\le t/a \le 10$) for the $24^3$ ($32^3$) ensembles.
The obtained values of $|f_S(q_{\rm max}^2)|$, which are naturally renormalized in the
double ratio, are summarized together with the values of $q_{\rm max}^2$ in Table~\ref{Tab:f0atqmax}.

Here, we note that the absolute value of the renormalized $f_S(q_{\rm max}^2)$
is exactly unity in the exact $SU(3)$ limit, where $f_S(q_{\rm max}^2)$
becomes $f_1^{\rm SU(3)}(0)=-1$ $(+1)$ for the $\StoN$ ($\XtoS$) beta decay, which 
is associated with the $SU(3)$ Clebsch-Gordan coefficients.
Thus, the deviation from unity in $|f_S^{\rm ren}(q_{\rm max}^2)|$ is 
attributed to three types of the $SU(3)$-breaking effect: (1) the recoil
correction ($q_{\rm max}^2\neq 0$) stemming from the mass difference
between the initial ($B_1$) and final ($B_2$) states, (2) the presence
of the second-class form factor $f_3(q^2)$, and (3) the deviation 
from its $SU(3)$ symmetric value $f_1^{\rm SU(3)}(0)$.
Indeed, our main target is to measure the third one.
In the next subsection,  we will thus evaluate the scalar form factor at $q^2=0$, $f_S(0)$, which
is identical to $f_1(0)$, in order to separate the third effect from the others.

\subsection{Interpolation to zero momentum transfer}
\label{Sec:Sec3_momfit}

The scalar form factor $f_S(q^2)$ at $q^2>0$, where the 
three-momentum transfer is finite ($|{\bm q}|\neq 0$), can be evaluated
through Eq.~(\ref{Eq:f_S}) with the three-point correlation functions
for both the time and space components of the vector current $V_{\alpha}$.
We use the four lowest nonzero momenta: ${\bm q}=2\pi/L \times (1,0,0)$,
$(1,1,0)$, $(1,1,1)$, and $(2,0,0)$, corresponding to a $q^2$ range 
from about 0.2 to 0.8 ${\rm GeV}^2$ in both $24^3$ and $32^3$ ensembles. 

Recall that the time-reversal process $B_2\rightarrow B_1$ provides
different $q^2$ points in comparison to that of $B_1 \rightarrow B_2$
even with the same nonzero three-momentum transfer ${\bm q}^2$ 
if the rest masses of the initial and final states are different. 
In this study, we then calculate both $B_1 \rightarrow B_2$ and $B_2 \rightarrow B_1$
processes in both $\StoN$ and $\XtoS$ beta-decay channels. 
Therefore, the four ${\bm q}^2$ calculations give eight data points 
of $f_S(q^2)$ in the range of $q^2>0$. 
We then can make the $q^2$ interpolation of $f_S(q^2)$ to $q^2=0$ by the values of 
$f_S(q^2)$ at $q^2>0$ together with the precisely measured value of $f_S(q^2)$ at $q^2=
q_{\rm max}^2<0$ from the double ratio as described in an earlier subsection.

In the $q^2$ interpolation, either a monopole form ($c_0/(1+c_2 q^2)$)
or the quadratic form ($c_0+c_2\cdot q^2$) have been adopted 
in the previous studies~\cite{{Guadagnoli:2006gj},{Sasaki:2008ha},{Sasaki:2012ne}}. 
However, the fitting form ans\"atz may tend to 
constrain the interpolation and introduce a model dependence into the final result
of the vector coupling $f_1(0)$. 
In order to reduce systematic errors associated
with an interpolation of the form factor in momentum transfer, 
we use the model-independent $z$ expansion 
method~\cite{{Boyd:1995cf},{Hill:2010yb}} in this study. 

Suppose that the form factor $f_S(q^2)$ is analytic on the complex plane of $q^2$ 
outside a branch cut running along the negative real axis ($q^2<0$). 
The $z$ expansion (denoted as z-Exp) makes use of a conformal mapping from $q^2$
to a new variable $z$~\cite{{Boyd:1995cf},{Hill:2010yb}}:
%
%
\be
z(q^2)=\frac{\sqrt{t_{\rm cut}+q^2}-\sqrt{t_{\rm cut}}}
{\sqrt{t_{\rm cut}+q^2}+\sqrt{t_{\rm cut}}},
\ee
where the branch point $t_{\rm cut}=(M_\pi+M_K)^2$ is associated with the $K\pi$ threshold energy
for the strangeness-changing weak decays. 
This transformation makes the analytic domain mapped 
inside a unit-circle $|z|<1$. The region where the data exist ($q^2\ge q_{\rm max}^2> -t_{\rm cut}$) is assured to be inside a circular region of analyticity~\cite{{Boyd:1995cf},{Hill:2010yb}}. 

The form factor $f_S(z)$ can be thus described 
by a convergent Taylor series in terms of $z$.
We therefore adopt the following fitting form 
%
%
\be
f_S(q^2)=\sum_{k=0}^{k_{\rm max}}a_k z(q^2)^k,
\ee
where $k_{\rm max}$ truncates an infinite series expansion in $z$. 
For a model-independent fit, $k_{\rm max}$ must ensure that terms $a_k z^k$ become numerically 
negligible for $k>k_{\rm max}$. 

In principle, there is an appropriate choice of $k_{\rm max}$ 
since $|a_{k}/a_{k-1}|<1$ is expected for sufficiently large $k$.
Recall that the range of possible values of $k_{\rm max}$, is limited 
by the condition $k_{\rm max}\le 7$ due to the limited nine data points of $f_S(q^2)$ in this study. 
In order to assess the stability of the fit results with a given $k_{\rm max}$, we plot the ratios of 
$|a_{k}/a_{k-1}|$, which are determined by fitting all of the nine data points using the z-Exp form 
with $k_{\rm max}=7$ in Fig.~\ref{fig:svd_coeffs}. 
As shown in Fig.~\ref{fig:svd_coeffs}, the ratios of $|a_{k}/a_{k-1}|$ reach 
a convergence value less than unity at $k\approx 3$. This implies that the z-Exp 
method gives a rapid convergence series which makes {\it a model independent fit}.

In Table~\ref{Tab:MomInterpolation}, we compile 
the results of $\tilde{f}_1(0)=f_1(0)/f^{\rm SU(3)}_1(0)$ obtained from 
the $q^2$-interpolation of $|f_S(q^2)|$ using the z-Exp fits with various choices of $k_{\rm max}$. 
Table~\ref{Tab:MomInterpolation} also contains the results given by the monopole and quadratic fits 
for comparison. 

\begin{figure*}[ht]
  \centering
  \includegraphics*[width=.48\textwidth]{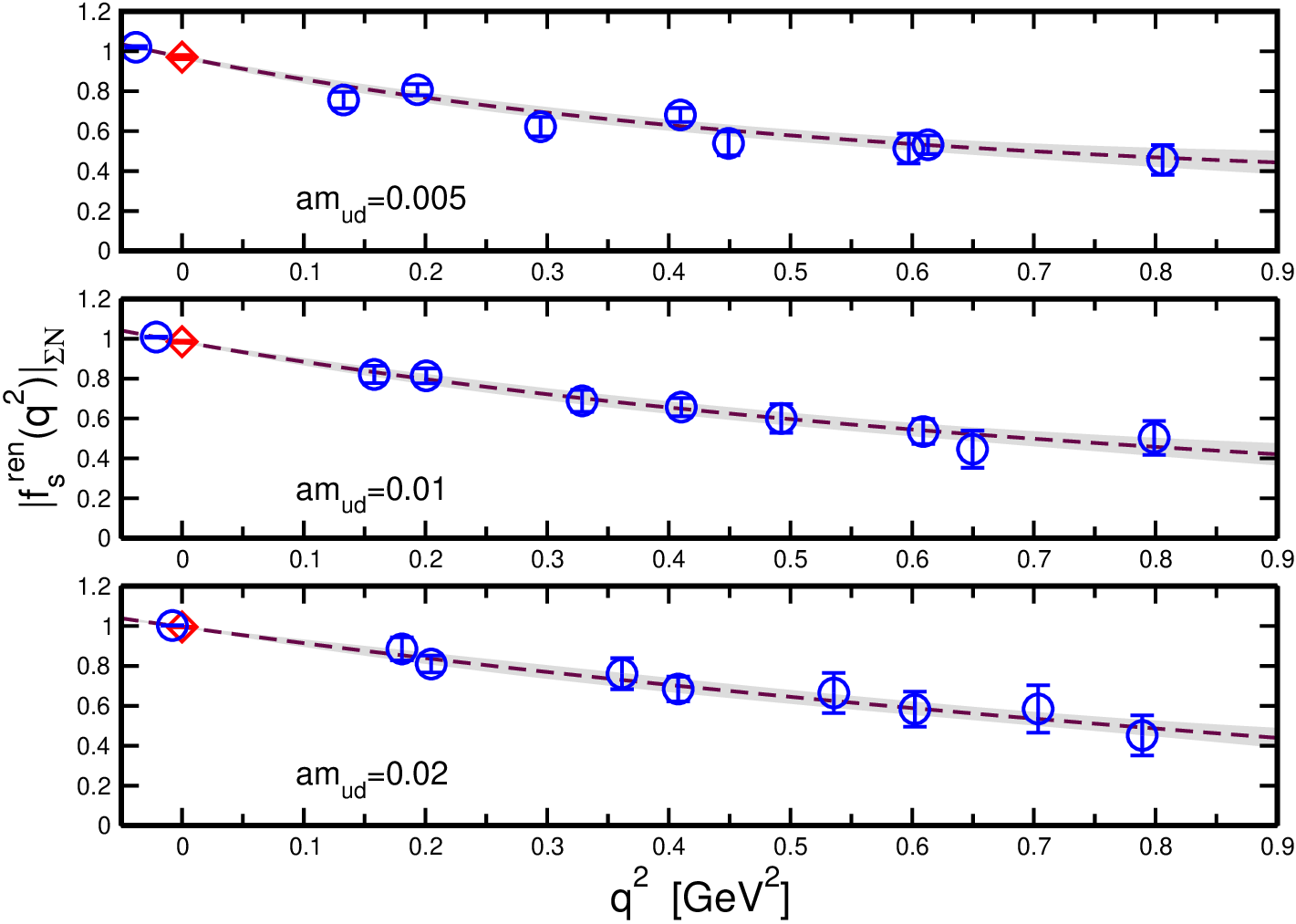} 
  \includegraphics*[width=.48\textwidth]{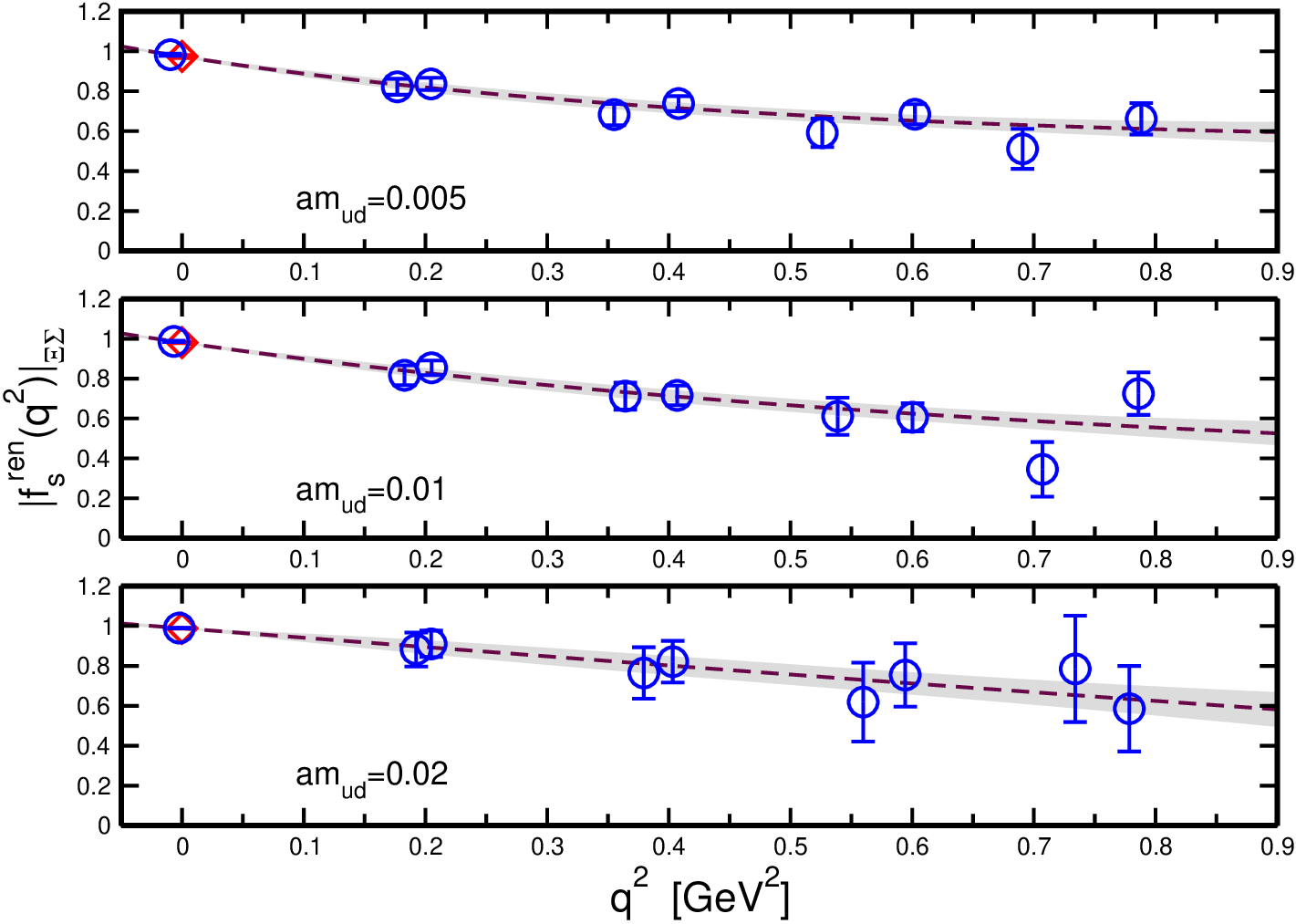} 
\caption{Interpolation of $|f_S(q^2)|$ to $q^2=0$ for the $24^3$ ensembles ($\beta=2.13$).
  The left (right) panel is for the $\StoN$ ($\XtoS$) beta decay at $am_{ud}=0.005$ (upper),
  0.01 (middle), and 0.02 (right). Open circles are the renormalized value of $|f_S(q^2)|$
  at the simulated $q^2$. A dashed curve in each plot is the fitting result by using the z-Exp method,
  while the open diamond represents the interpolated value to $q^2=0$.}
  \label{fig:f0mom_b213}
  \end{figure*}

\begin{figure*}[ht]
  \centering
  \includegraphics*[width=.48\textwidth]{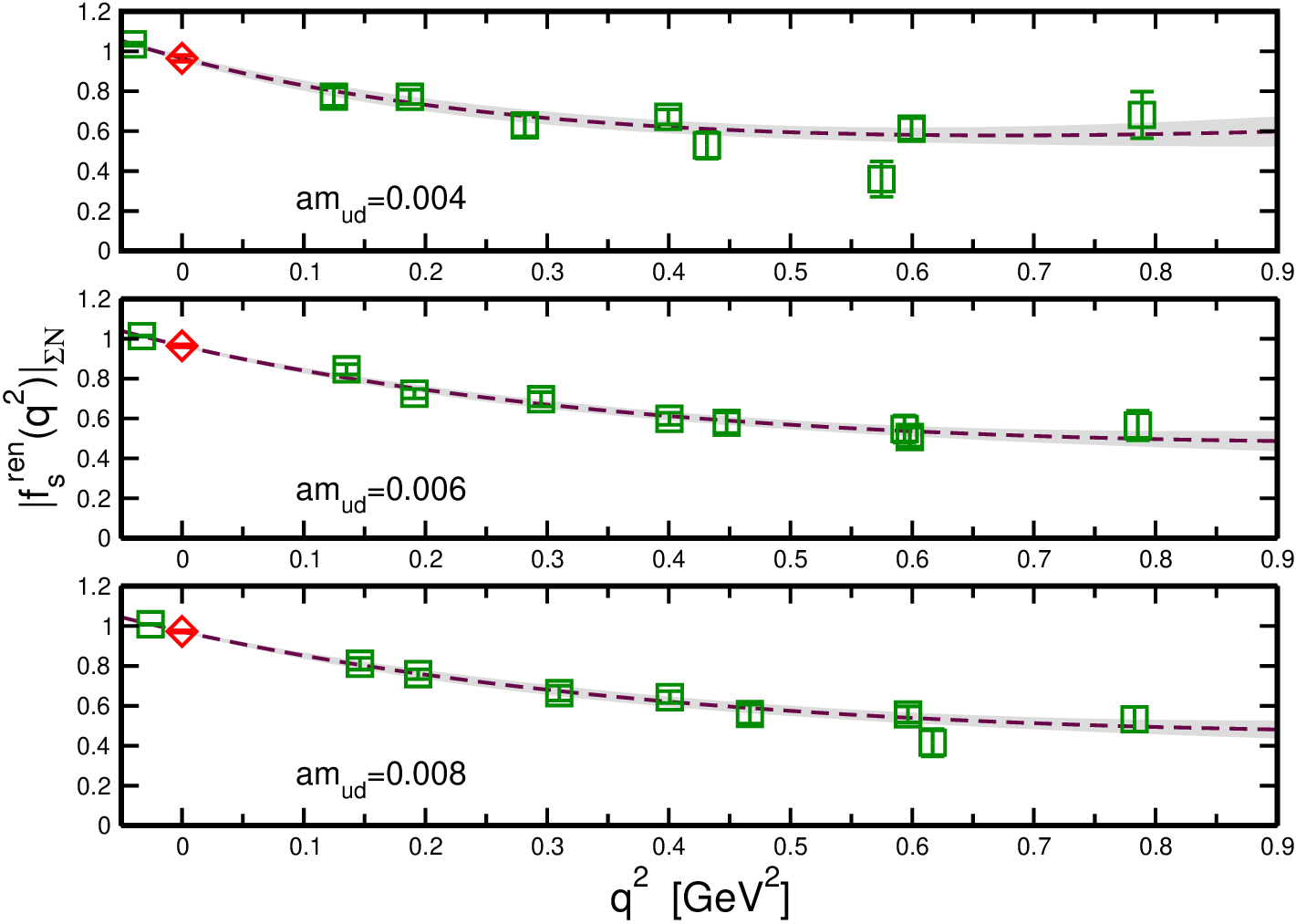} 
  \includegraphics*[width=.48\textwidth]{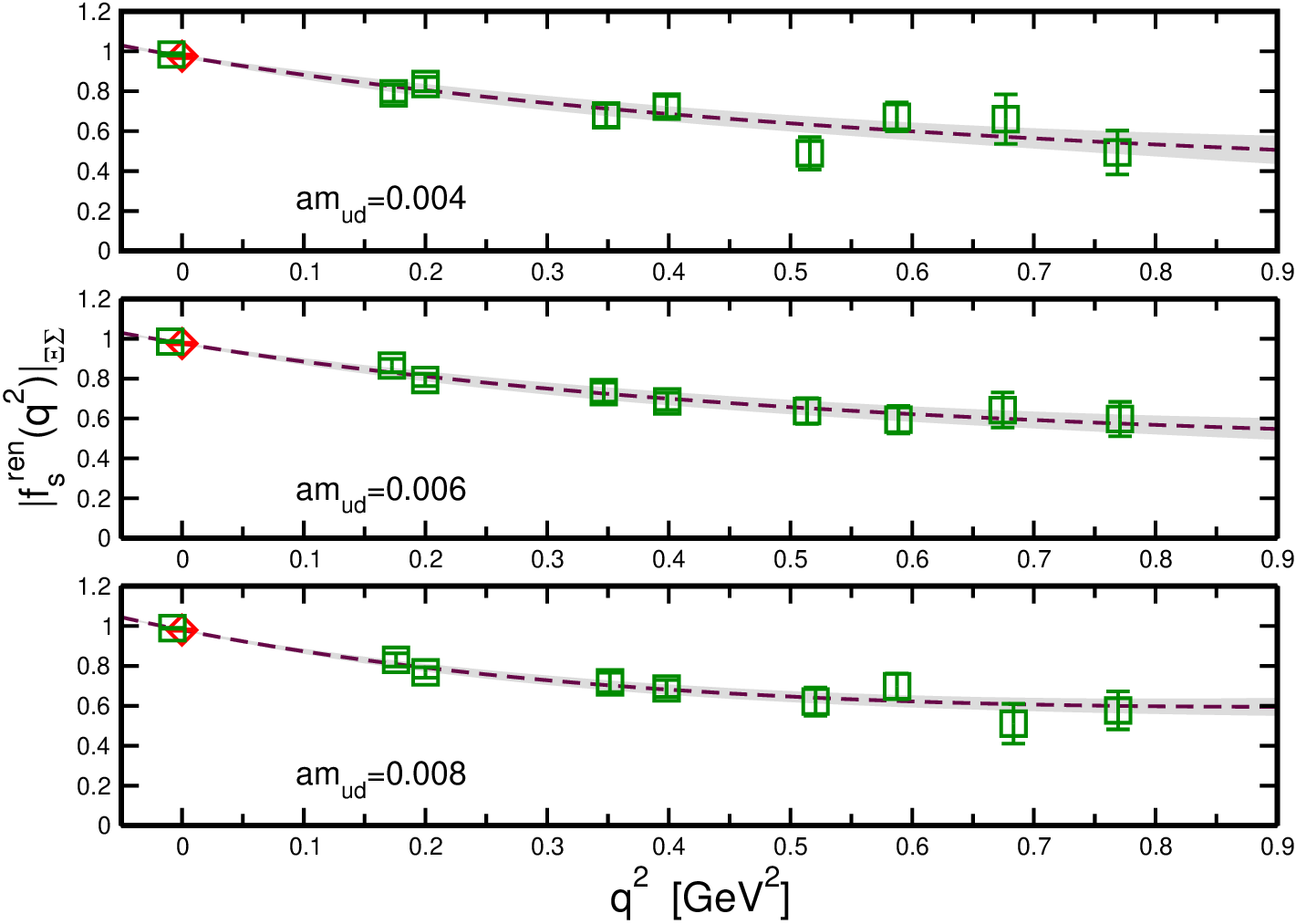} 
  \caption{Interpolation of $|f_S(q^2)|$ to $q^2=0$ for the $32^3$ ensembles ($\beta=2.25$).
  The left (right) panel is for the $\StoN$ ($\XtoS$) beta decay at $am_{ud}=0.004$ (upper),
  0.006 (middle), and 0.008 (right). Open squares are the renormalized value of $|f_S(q^2)|$
  at the simulated $q^2$. A dashed curve in each plot is the fitting result by using the z-Exp method, 
  while the open diamond represents the interpolated value to $q^2=0$.}
  \label{fig:f0mom_b225}
  \end{figure*}

First of all, as expected in Fig.~\ref{fig:svd_coeffs}, the interpolated value of $\tilde{f}_1(0)$ 
is not sensitive to the choice of $k_{\rm max}$ in the z-Exp fits. Furthermore, the inclusion 
of the higher powers in $z$ does not reduce $\chi^2/{\rm dof}$ significantly. 
For these reasons, we hereafter choose $k_{\rm max}=3$ in the z-Exp method. 
Examples of the $q^2$-interpolation of $|f_S(q^2_{\rm max})|$ given by the z-Exp method
are shown in Figs.~\ref{fig:f0mom_b213} and \ref{fig:f0mom_b225}.
As can be seen from those figures, $f_1(0)$ can be determined by a very short interpolation 
from $q_{\rm max}^2$, where we have very accurate data $|f_S(q^2_{\rm max})|$ 
from the double ratio (\ref{Eq:DoubleRatio}). This is the reason why the choice of the $q^2$-interpolation form 
does not much affect the interpolated value $f_1(0)$ significantly.

%
%
%

%
%
%

%
%
\begin{table*}[!t]
\caption{
Results for $\tilde{f}_1(0)=f_1(0)/f^{\rm SU(3)}_1(0)$, where $f^{\rm SU(3)}_1(0)=+1$ for the $\XtoS$ beta decay
and $f^{\rm SU(3)}_1(0)=-1$ for the $\StoN$ beta decay, by using various $q^2$-interpolation forms.}
\label{Tab:MomInterpolation}
\begin{ruledtabular}
\begin{tabular}{cl l c l c l c }
\hline
 $\beta=2.13$&&  \multicolumn{2}{@{}c@{}}{$am_{ud}=0.005$}
&  \multicolumn{2}{@{}c@{}}{$am_{ud}=0.010$}&  \multicolumn{2}{@{}c@{}}{$am_{ud}=0.020$}\cr
Decay &$q^2$ interpolation
& \multicolumn{1}{c}{$\tilde{f}_1(0)$} & $\chi^2/{\rm dof}$
& \multicolumn{1}{c}{$\tilde{f}_1(0)$} &  $\chi^2/{\rm dof}$ 
& \multicolumn{1}{c}{$\tilde{f}_1(0)$} &  $\chi^2/{\rm dof}$ \cr
\hline
& monopole fit&
0.9713(83)& 9.54/7 &0.9820(45)& 1.57/7 &0.9943(16) & 1.96/7\cr
& quadratic fit&
0.9761(81)& 11.37/6 &0.9863(44)& 1.20/6  &0.9959(16)& 1.55/6 \cr
$\StoN$ &   z-Exp fit ($k_{\rm max}=2$) &
0.9711(93) & 10.65/6 &0.9849(50)& 1.34/6 &0.9956(18) & 1.51/6\cr
&   z-Exp fit ($k_{\rm max}=3$) &
0.9713(93)& 10.66/5 &0.9849(50)& 1.33/5 &0.9956(18) & 1.51/5\cr
&   z-Exp fit ($k_{\rm max}=7$) &
0.9713(92)& 10.66/1 &0.9849(50)& 1.33/1 &0.9956(18) & 1.51/1\cr
\hline
 & monopole fit & 
0.9753(56)&  5.47/7 &0.9811(32)& 5.43/7 & 0.9885(13)  & 0.99/7\cr
& quadratic fit &
0.9751(57)&  5.25/6 & 0.9811(32)& 6.42/6 & 0.9887(13) & 1.05/6 \cr
$\XtoS$ &   z-Exp fit ($k_{\rm max}=2$) &
0.9742(59) & 5.49/6 &0.9808(33) & 6.70/6 &0.9887(14)& 1.04/6  \cr
&   z-Exp fit ($k_{\rm max}=3$) &
0.9742(59)& 5.48/5 &0.9808(33)& 6.69/5 &0.9887(14) & 1.04/5 \cr
&   z-Exp fit ($k_{\rm max}=7$) &
0.9742(59)& 5.48/1 &0.9808(33)& 6.69/1 &0.9887(14) & 1.04/1\cr
\hline
\hline
$\beta=2.25$&&  \multicolumn{2}{@{}c@{}}{$am_{ud}=0.004$}
&  \multicolumn{2}{@{}c@{}}{$am_{ud}=0.006$}&  \multicolumn{2}{@{}c@{}}{$am_{ud}=0.008$}\cr
Decay &$q^2$ interpolation
& \multicolumn{1}{c}{$\tilde{f}_1(0)$} & $\chi^2/{\rm dof}$
& \multicolumn{1}{c}{$\tilde{f}_1(0)$} &  $\chi^2/{\rm dof}$ 
& \multicolumn{1}{c}{$\tilde{f}_1(0)$} &  $\chi^2/{\rm dof}$\cr
\hline
 & monopole fit&
0.9819(118) & 15.40/7 & 0.9663(66) & 8.08/7 &0.9733(51) & 4.48/7 
\cr
& quadratic fit&
0.9754(128) & 12.32/6 & 0.9690(62) & 4.76/6 &0.9761(45) & 4.57/6
\cr
$\StoN$&   z-Exp fit ($k_{\rm max}=2$) &
0.9650(149) & 12.60/6 & 0.9641(71)& 6.03/6 & 0.9725(53) & 4.70/6
\cr
&   z-Exp fit ($k_{\rm max}=3$) &
0.9654(148)& 12.58/5 & 0.9643(71)& 5.98/5 & 0.9726(53) &  4.68/5
\cr
&   z-Exp fit ($k_{\rm max}=7$) &
0.9655(148) &12.57/1 &0.9643(71)& 5.98/1 & 0.9726(52) & 4.68/1
\cr
\hline
 & monopole fit&
0.9760(70) & 5.69/7 & 0.9766(38) & 2.49/7 & 0.9812(29) & 5.71/7 
\cr
& quadratic fit&
0.9762(69) & 7.02/6 & 0.9769(38) & 1.92/6 & 0.9810(29) & 4.46/6
\cr
$\XtoS$ &   z-Exp fit ($k_{\rm max}=2$) &
0.9755(72) & 6.98/6 &0.9761(40) & 2.24/6 &0.9798(30) & 3.80/6
\cr
&  z-Exp fit ($k_{\rm max}=3$) &
0.9755(72)& 6.98/5 & 0.9761(40)& 2.23/5 &0.9799(30)& 3.82/5
\cr
&  z-Exp fit ($k_{\rm max}=7$) &
0.9755(72)& 6.98/1 & 0.9761(40)& 2.22/1 &0.9799(30) & 3.82/1
\cr
\hline
\end{tabular}
\end{ruledtabular}
\end{table*}
%

\subsection{Chiral and continuum extrapolation of $f_1(0)$}
\label{Sec:Sec3_chiral_cont}

We next perform the chiral extrapolation of $f_1(0)$ in order 
to estimate $f_1(0)$ at the physical point. In our previous work~\cite{Sasaki:2012ne},
we adopt a global fit of the data on $\tilde{f}_1(0)=f_1(0)/f_1^{\rm SU(3)}(0)$ 
as multiple functions of $M_K^2-M_\pi^2$ and $M_K^2+M_\pi^2$ as
%
%
\be
\tilde{f}_1(0)=C_0+(C_1+C_2\cdot(M_K^2+M_\pi^2))\cdot(M_K^2-M_\pi^2)^2,
\label{Eq:GlobalFit_type1}
\ee
whose form (denoted as Type 1) is motivated by 
the AGT~\cite{Sasaki:2008ha}.
Our simulations on both $24^3$ and $32^3$ ensembles are performed with a strange quark 
mass slightly heavier than the physical mass~\cite{{Allton:2008pn},{Aoki:2010dy}}.
Therefore, the third term that is proportional to $M_K^2+M_\pi^2$ can manage to 
compensate for a small difference in the simulated and physical strange-quark masses 
in an {\it a posteriori} way.  

\begin{figure*}[ht]
\centering
\includegraphics[width=.48\textwidth,clip]{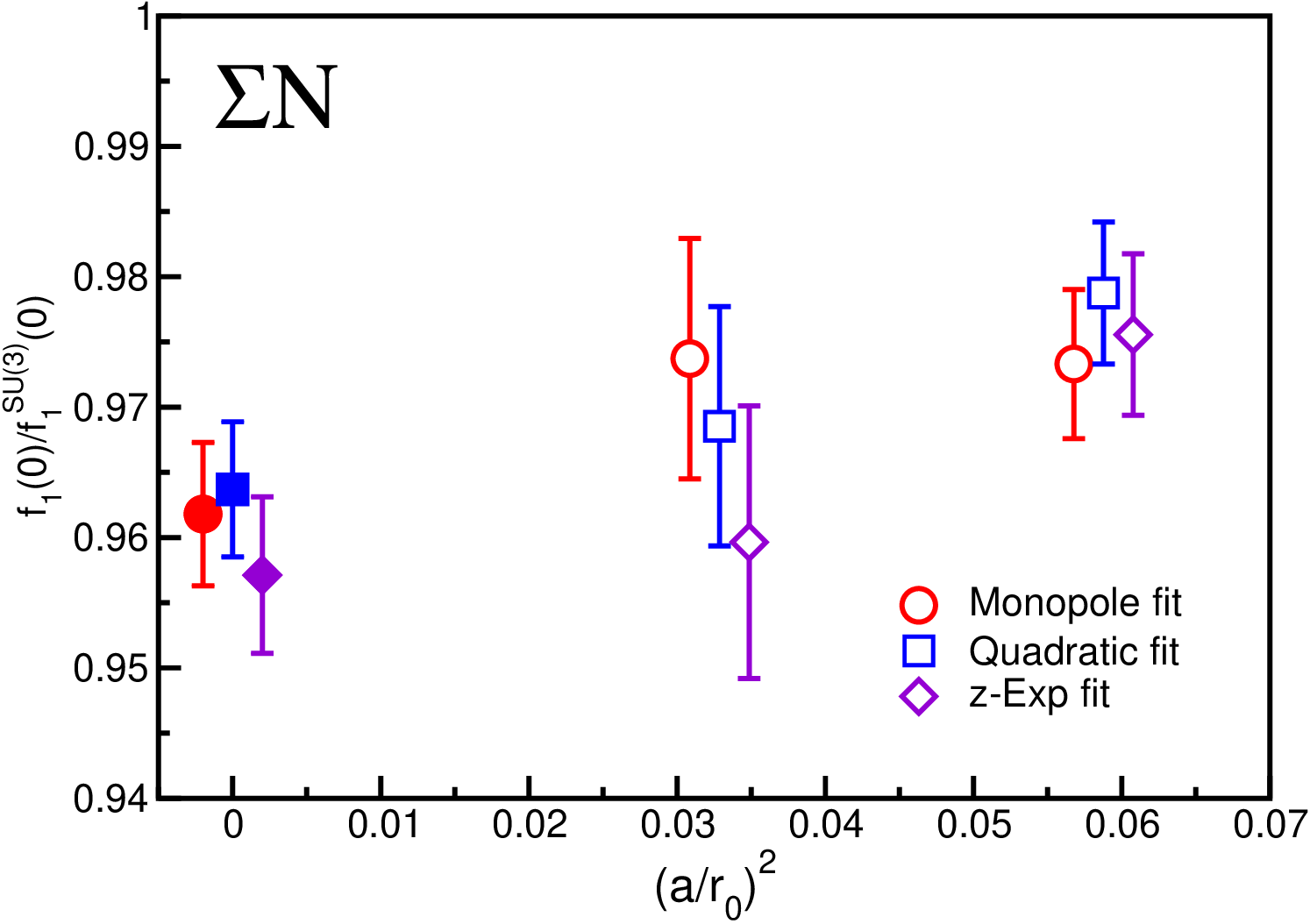}
\includegraphics[width=.48\textwidth,clip]{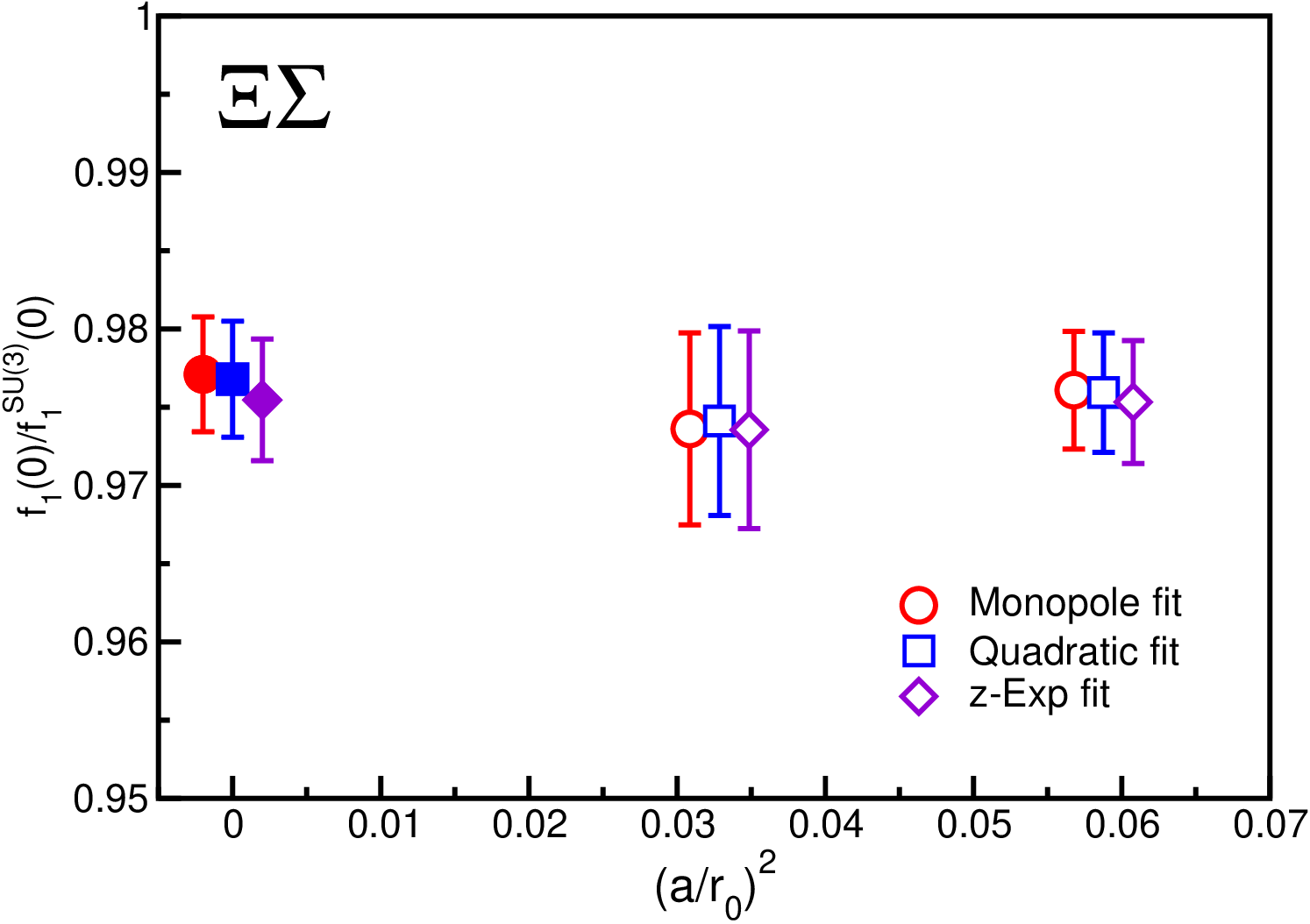}
\caption{The scaling behavior of $\tilde{f}_1(0)$ versus $(a/r_0)^2$. 
The left (right) panel is for $\StoN$ ($\XtoS$) beta decay. 
Open symbols are obtained from separate chiral extrapolations on
the $24^3$ and the $32^3$ data sets. 
The continuum extrapolated values, which are determined 
by the combined continuum-chiral fits (Type 3), are also included 
as filled symbols for comparison.  Recall that the filled symbols do not correspond to the results 
given by a naive linear extrapolation on the open symbols. 
The values of the Sommer scale $r_0/a$ are taken from Ref.~\cite{Aoki:2010dy}. 
}\label{Fig:adep_f1}
\end{figure*}

We first test the global fit on the results from the $24^3$ and $32^3$ ensembles separately. 
In Fig.~\ref{Fig:adep_f1}, we plot the extrapolated values of $\tilde{f}_1(0)$ at the physical point
(open symbols) as a function of $(a/r_0)^2$ where $r_0$ denotes the Sommer scale~\cite{Sommer:1993ce}. 
Different symbols, which are consistent with each other within their errors, 
represent results from three different interpolations: monopole, quadratic and z-Exp fits. 
It is found that there is no significant scaling violation due to the lattice discretization 
in the vector couplings for both $\StoN$ and $\XtoS$ beta decays.

We then perform a combined global-fit of both $24^3$ and $32^3$ lattice data on 
$\tilde{f}_1(0)$ determined from the z-Exp fits by using the Type 1 formula [Eq.~(\ref{Eq:GlobalFit_type1})]
ignoring possible discretization errors. Fit results (Type 1 fit) are tabulated in Table~\ref{Tab:GlobalFitResults}. 
We then get the vector coupling $f_1(0)$ at the physical point as 
\begin{equation}
f_1^{\StoN}(0)=-0.9662(43),\quad f_1^{\XtoS}(0)=+0.9742(28),
\end{equation}
where the quoted errors are only statistical. 
The inclusion of the new ensembles in our combined global-fit 
leads to a reduction of the statistical error at the physical point compared 
to our earlier work~\cite{Sasaki:2012ne}, which is performed only 
on the $24^3$ ensembles with less number of measurements.

Here, we recall that the value of $C_0$ is supposed to be unity since the vector current 
conservation at $M_K=M_\pi$, while $C_0$ obtained from 
the global fitting form~(\ref{Eq:GlobalFit_type1}) is slightly off the unity 
beyond the statistical uncertainty as listed in Table~\ref{Tab:GlobalFitResults}. 
The lattice discretization error could be an origin of its slight deviation from the unity.

%
%
%

%
%
\begin{table*}[!t]
\caption{The coefficients of three types of the combined global fit to
all data of $\tilde{f}_1(0)$ calculated on the $24^3$ and $32^3$ ensembles.
The renormalized values of $f_1(0)$ are evaluated at each simulated quark mass
by the $q^2$-interpolation with the z-Exp method. 
}
\label{Tab:GlobalFitResults}
\begin{ruledtabular}
\begin{tabular}{c c ll lr c l}
Decay & Global fit 
& \multicolumn{1}{c}{$C_0$} 
& \multicolumn{1}{c}{$C_1$ [$({\rm GeV})^{-4}$]}
& \multicolumn{1}{c}{$C_2$ [$({\rm GeV})^{-6}$]}
& \multicolumn{1}{c}{$C_3$ [$({\rm GeV})^{2}$]} 
& $\chi^2/{\rm dof}$ & \multicolumn{1}{c}{$f_1(0)/f^{\rm SU(3)}_1(0)$}\cr
\hline
$\StoN$ & Type 1 & 1.0131(42) & $-0.844(140)$ &$-0.232(39)$& \multicolumn{1}{c}{N/A} 
& 1.69 & 0.9662(43)\cr
& Type 2 & 0.9795(180) & $-0.587(191)$ & $-0.182(61)$ & $0.086(43)$ & 0.34 & 0.9466(109)\cr
&Type 3 & 1.0 (fixed) & $-0.757(106)$& $-0.270(38)$& $0.038(10)$ & 0.74 & 0.9571(60)\cr
\hline
$\XtoS$ &Type 1 & 0.9972(30) & $-0.416(96)$ & $-0.106(26)$ & \multicolumn{1}{c}{N/A} 
& 0.09 & 0.9742(28)\cr
&Type 2 & 0.9943(98) & $-0.386(122)$ & $-0.116(38)$& $0.008(22)$ & 0.09 & 0.9727(57)\cr
&Type 3 & 1.0 (fixed) & $-0.433(69)$ & $-0.155(24)$ &$-0.005(7)$ & 0.16 & 0.9755(39)\cr
\hline
\end{tabular}
\end{ruledtabular}
\end{table*}

To take into account the lattice discretization corrections into the fitting form ans\"atz, 
let us introduce the second type of the global fit (denoted as Type 2), which is given by
\begin{multline}
\tilde{f}_{1}(0)=\left( C_0 + C_3 a^2 \right)\\
+ \left( C_1 +  C_2 \cdot (M_{K}^2+M_{\pi}^2 )\right)\cdot (M_{K}^2-M_{\pi}^2)^2 ,
\label{Eq:GlobalFit_type2}
\end{multline}
where $C_3$ coefficient takes into account the lattice discretization error on each data of $f_1(0)$ calculated
at two different lattice spacings as the leading-order term. 
In fact, an inclusion of the $a^2$ correction term in the global fit formula 
certainly cures the unity condition on $C_0$ albeit with larger statistical uncertainties on each coefficient
as shown in Table~\ref{Tab:GlobalFitResults}.
Although the size of $C_3$ is very small compared to other coefficients, its inclusion in the fitting 
ans\"atz is statistically relevant especially for $\StoN$ decay data. 

Finally, we set $C_0=1$ as a theoretical constraint associated to the $SU(3)$ symmetric value 
in continuum and then propose the third fitting formula (denoted as Type 3)
\begin{multline}
\tilde{f}_{1}(0)=\left(1 + C_3 a^2\right) \\
+ \left( C_1 +  C_2 \cdot (M_{K}^2+M_{\pi}^2 )\right)\cdot (M_{K}^2-M_{\pi}^2)^2 ,
\label{Eq:GlobalFit_type3}
\end{multline}
which gives the better statistical uncertainties on all coefficients, whose values are consistent with 
the fit results by the Type 2 formula [Eq.~(\ref{Eq:GlobalFit_type2})]
as summarized in Table~\ref{Tab:GlobalFitResults}. 
We therefore choose the Type 3 formula for evaluating the final result of $\tilde{f}_1(0)$
at the physical point. 

\begin{figure*}[ht]
\centering
\includegraphics[width=.48\textwidth,clip]{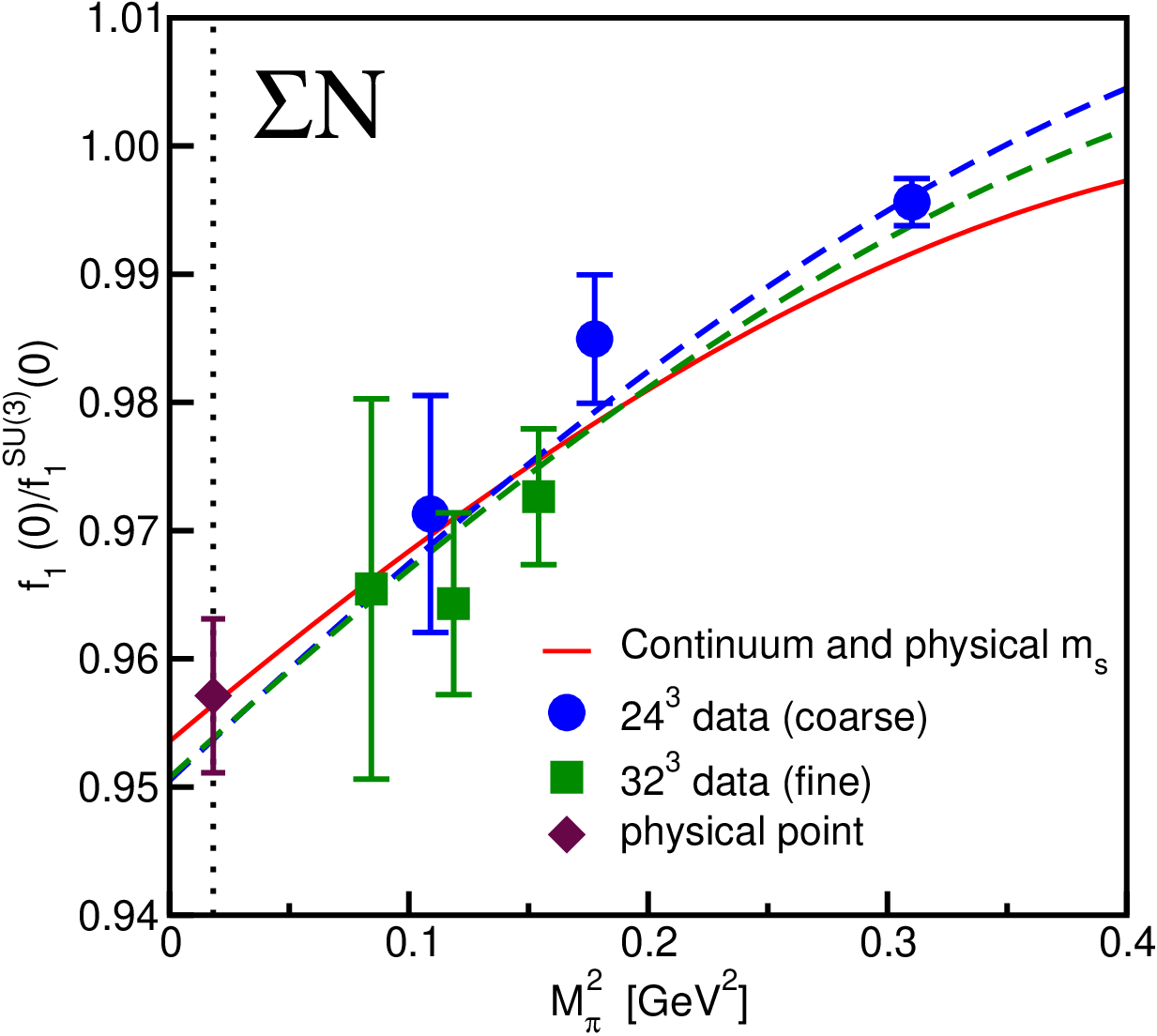}
\includegraphics[width=.48\textwidth,clip]{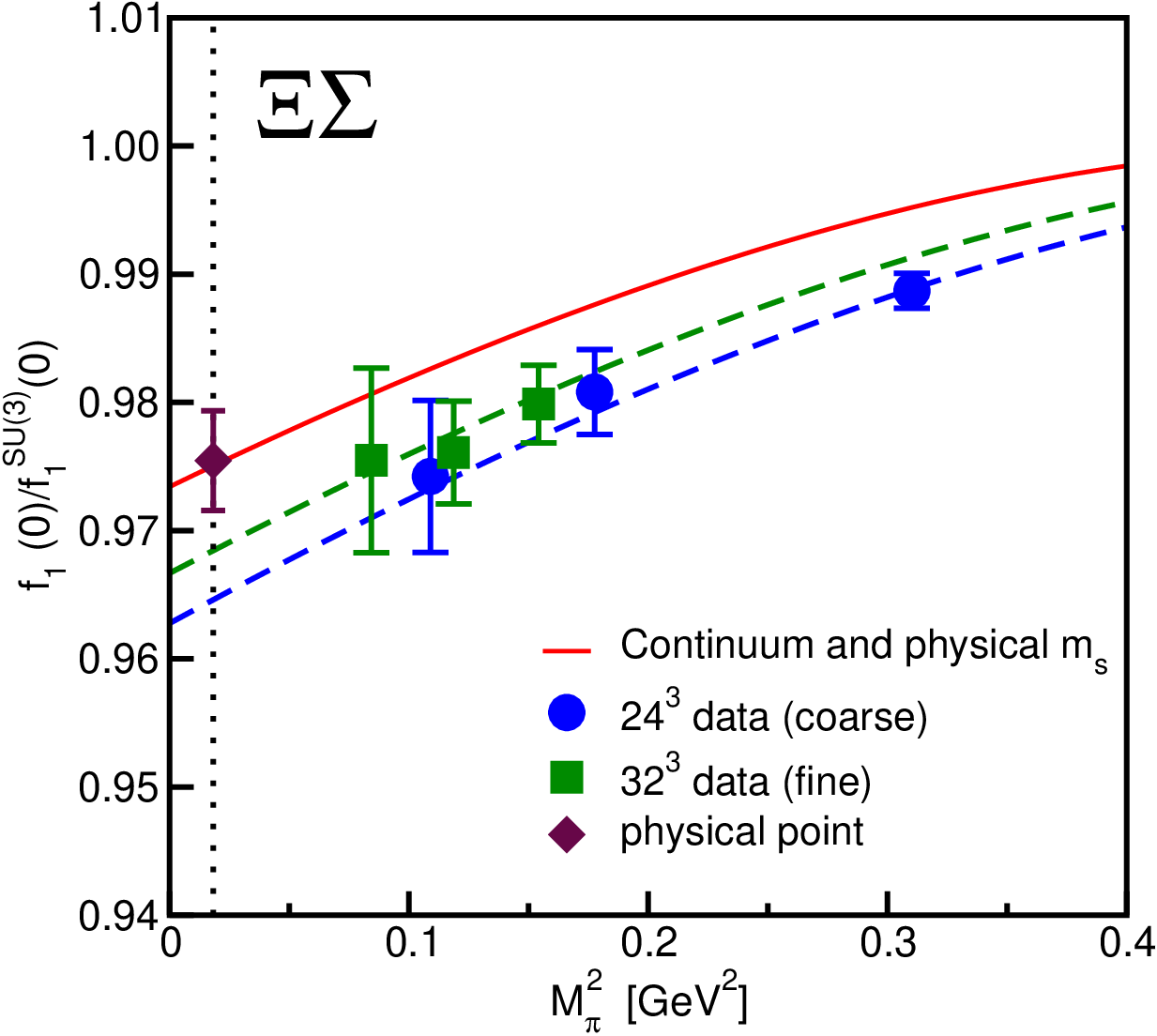}
\caption{Chiral and continuum extrapolation of $\tilde{f}_1(0)$ for the $\StoN$ (left panel) 
and $\XtoS$ (right panel) beta decays using Eq.~(\ref{Eq:GlobalFit_type3}) (Type 3 fit). 
In each panel, the filled diamonds denote the continuum value
of $\tilde{f}_1(0)$ at the physical point. 
}\label{Fig:ChiraExtraRaw_wasq_su3fix}
\end{figure*}

In Fig.~\ref{Fig:ChiraExtraRaw_wasq_su3fix}, we 
plot the results of $\tilde{f}_1(0)$ for the $\StoN$ (left panel) and $\XtoS$ (right panel)
beta decays as a function of $M_{\pi}^2$ together with
the continuum value of $\tilde{f}_1(0)$ at the physical point (diamond symbol), 
that is determined through the combined global-fit of both $24^3$ (circle symbols) and $32^3$ 
lattice data (squared symbols) with the Type 3 formula (Eq.~(\ref{Eq:GlobalFit_type3})). 
In each panel, fitting curves indicated by dashed curves represent the simultaneous fitting
results on each data set calculated at all simulated quark masses.
The solid curve corresponds to the continuum results given at the 
physical strange quark mass.

We then get the continuum values of the vector coupling $f_1(0)$ at the physical point as 
%
%
\begin{equation}
f_1^{\StoN}(0)=-0.9571(60),\quad f_1^{\XtoS}(0)=+0.9755(39),
\end{equation}
where the systematic uncertainties due to the lattice discretization error are also
included in the quoted errors as well as the statistical one. 
These values are shown as filled diamond symbols
in Fig.~\ref{Fig:adep_f1}. The filled circle and squared symbols are the extrapolated results
from data of $f_1(0)$ given by the different $q^2$ interpolations. 
Although the extrapolated value at the physical point in the continuum does not
significantly depend on which type of $q^2$ interpolation as shown in Table~\ref{Tab:FinalResults}, 
we simply quote the systematic uncertainties
due to $q^2$ interpolation as the maximum difference among three types of $q^2$ interpolations.
As for the systematic uncertainty of the chiral extrapolation, we read off a difference in the extrapolated
values with and without the $C_2$ coefficient, which is associated with corrections beyond the AGT, 
in the Type 3 formula.  Hence our final results are
%
%
\be
f_1(0)= \left\{
\begin{array}{ll}
-0.9571(60)_{\rm stat}(66)_{q^2}(37)_{\chi}(24)_{\rm scale}& [\StoN]\cr
+0.9755(39)_{\rm stat}(16)_{q^2}(21)_{\chi}(24)_{\rm scale} & [\XtoS],
\end{array}
\right.
\ee
where the first error is statistical, and the second, third and fourth are 
estimates of the systematic errors due to our
choice of $q^2$-interpolation, the reliability of the extrapolation to the
physical point, and the uncertainty of the scale parameter. 

%
%
\begin{table}[ht]
\caption{
Results for the continuum value of $\tilde{f}_1(0)=f_1(0)/f^{\rm SU(3)}_1(0)$
at the physical point for the $\StoN$ and $\XtoS$ beta decays. }
\label{Tab:FinalResults}
\begin{ruledtabular}
\begin{tabular}{cc lll}
\hline
& & \multicolumn{3}{@{}c@{}}{combined global fit} \cr
 Decay & $q^2$ interpolation
 & \multicolumn{1}{c}{Type 1}
 & \multicolumn{1}{c}{Type 2} 
 & \multicolumn{1}{c}{Type 3}\cr
\hline 
& monopole fit &
0.9683(38) & 0.9549(100) & 0.9618(55)
\cr
$\StoN$ & quadratic fit&  
0.9711(37) & 0.9546(94) & 0.9637(52)
\cr
&  z-Exp fit & 
0.9662(43) & 0.9466(109) & 0.9571(60)
\cr
\hline
 & monopole fit &
0.9753(26) & 0.9748(54) & 0.9771(37)
\cr
$\XtoS$& quadratic fit & 
0.9752(26) & 0.9745(54) & 0.9768(37)
\cr
& z-Exp fit  & 
0.9742(28) & 0.9727(57) & 0.9755(39)
\cr
\hline
\end{tabular}
\end{ruledtabular}
\end{table}
%

%
%

The remaining source of systematic uncertainty is due to 
the finite-volume used in lattice simulation, where
the physical spatial extent is approximately 2.7 fm for
both $24^3$ and $32^3$ ensembles. 
The previous studies of the nucleon structure with the $24^3$ ensembles
reported that the nucleon vector form factor at low $q^2$ does not suffer much from 
the finite-volume effect though such effect may influence other nucleon form factors, especially 
the axial-vector one~\cite{{Yamazaki:2008py},{Yamazaki:2009zq}}.
Therefore, one may deduce that a lattice volume of $(2.7\;{\rm fm})^3$
used in our simulations is large enough to safely ignore finite volume corrections 
to the hyperon vector coupling in comparison to other systematic uncertainties. 

Adding all sources of error in quadrature, we obtain
%
%
\be
f_1(0)= \left\{
\begin{array}{ll}
-0.9571(99)_{\rm combined} & [\StoN]\cr
+0.9755(53)_{\rm combined} & [\XtoS],
\end{array}
\right.
\label{Eq:FinalResult_f1}
\ee
both of which reach an accuracy of about 1\% (or less). 
The $SU(3)$-breaking corrections
$\Delta f$ for two decays are also obtained as
%
%
\be
\Delta f = \left\{
\begin{array}{ll}
-0.0429(99) & [\StoN]\cr
-0.0245(53) & [\XtoS],
\end{array}
\right.
\label{Eq:Delta_f}
\ee
which are {\it both negative}. It is worth emphasizing that the signs of the $SU(3)$-breaking correction $\Delta f$
are consistent with what was reported in earlier lattice studies including both
quenched simulations~\cite{{Guadagnoli:2006gj},{Sasaki:2008ha}} and unquenched simulations~\cite{{Lin:2008rb},{Gockeler:2011se},{Sasaki:2012ne},{Shanahan:2015dka}}.  Furthermore, the sizes of $\Delta f$ 
for the $\StoN$ and $\XtoS$ beta decays are comparable to what was observed in the DWF calculations
of the $K_{l3}$ decays~\cite{Boyle:2007qe}. 
We however recall that the tendency of the $SU(3)$-breaking correction observed here
disagrees with predictions of the latest baryon chiral perturbation theory (ChPT) result up to ${\cal O}(p^4)$~\cite{{Geng:2009ik},{Geng:2014efa}} and the earlier large $N_c$ analysis~\cite{{Flores-Mendieta:1998ii},{FloresMendieta:2004sk}}.

In the baryon ChPT, the ${\cal O}(p^3)$ corrections are in general larger than the ${\cal O}(p^2)$ 
calculations leading often to a sign reversal 
of $\Delta f$~\cite{{Geng:2009ik},{Geng:2014efa},{Villadoro:2006nj},{Lacour:2007wm}}. 
There is clearly the convergence problem in the chiral expansion.
In fact, the leading corrections of ${\cal O}(p^2)$ to $f_1(0)$ are barely consistent with the lattice results of 
$\Delta f$~\cite{{Geng:2009ik},{Geng:2014efa},{Villadoro:2006nj},{Lacour:2007wm}}. 
On the other hand, the large $N_c$ analysis has received some criticism from Mateu and Pich~\cite{Mateu:2005wi}.
They pointed out that the large-$N_c$ fit including second-order $SU(3)$-breaking effects on $f_1(0)$ becomes
unreliable within the present experimental uncertainties. 

Recently, Flores-Mendieta and Goity have proposed a new framework of the chiral expansion,
that is consistent with the $1/N_c$ expansion of QCD~\cite{Flores-Mendieta:2014vaa}. 
They then provided the complete ${\cal O}(p^2)$ corrections to $f_1(0)$, 
which is consistent with the lattice results of $\Delta f$~\cite{Flores-Mendieta:2014vaa}. 
However, recall that the ${\cal O}(p^3)$ corrections, that expose some contradiction in 
other types of the baryon ChPT, have been not yet evaluated.

Next let us compare our results of $f_1(0)$ to experiments. 
Using the best estimate of $|V_{us}|=0.2254(8)$ with 
imposing CKM unitarity~\cite{Antonelli:2010yf},
we then predict the values 
%
%
\be
\begin{array}{lcl}
|V_{us}f_1(0)|_{\StoN}&=&0.2157(8)_{V_{us}}(22)_{f_1},\cr
|V_{us}f_1(0)|_{\XtoS}&=&0.2199(8)_{V_{us}}(12)_{f_1}
\end{array}
\ee
using our results given in Eq.~(\ref{Eq:FinalResult_f1}).
The first error comes from the error of $V_{us}$, and the second is the combined error of $f_1(0)$.
Although the latter decay is barely consistent with a single experimental result of 
$|V_{us}f_1(0)|_{\XtoS}=0.209(27)$~\cite{AlaviHarati:2001xk}, the former decay is slightly deviated from
the currently available experimental result of $|V_{us}f_1(0)|_{\StoN}=0.2282(49)$~\cite{Hsueh:1988ar}
and then reveals more than $2\sigma$ tension.

This discrepancy might be explained by the following reason. 
Through a polarized-$\Sigma^-$ beta-decay experiment, 
$g_1(0)/f_1(0)$ can be determined as a function of $g_2(0)/f_1(0)$~\cite{Cabibbo:2003cu}. 
This yields the constraint $g_1(0)/f_1(0)-0.133g_2(0)/f_1(0)=-0.327(20)$
for the $\StoN$ beta decay~\cite{Hsueh:1988ar}. 
Then, the conventional assumption $g_2(0)=0$ gives the final value of $g_1(0)/f_1(0)=-0.327(20)$,
that is used in the experimental analysis on $|V_{us}f_1(0)|_{\StoN}$ determined 
from the decay rate of Eq.~(\ref{Eq:DecayRate})~\cite{{Cabibbo:2003cu},{Hsueh:1988ar}}.
The assumption $g_2(0)=0$ is no longer valid without the exact $SU(3)$ 
flavor symmetry~\cite{Weinberg:1958ut}.
Therefore, a few $\sigma$ discrepancy may be associated with this assumption made when  
estimating the value of $g_1(0)/f_1(0)$. 
 
The value of $g_2(0)$ should be subject to the first order corrections of $SU(3)$ 
breaking, which are an order of 10-15\%. 
Indeed, non-zero values of $g_2(0)$ are reported as the size of the first order corrections from 
quenched lattice QCD
for both $\StoN$~\cite{Guadagnoli:2006gj} and $\XtoS$~\cite{Sasaki:2008ha} beta-decay channels. 
On the other hand, a test of the CKM unitarity through 
the first row relation $|V_{ud}|^2+|V_{us}|^2+|V_{ub}|^2=1$ reaches  
a sub-percent level accuracy using the value of $V_{us}$ given 
by the average of the $K_{l3}$ and $K_{\mu 2}$ determinations~\cite{PDG}.
Therefore, let us now use the CKM unitarity together with our theoretical estimate of $f_1(0)$
so as to read off $g_2(0)$ from the $\StoN$ beta-decay rate and the constraint  
$|g_1(0)/f_1(0)-0.133g_2(0)/f_1(0)|_{\StoN}=0.327(20)$ in experiments~\cite{Hsueh:1988ar}.
We thus estimate 
%
%
\be
g_2(0)= 0.57(20)\quad{\rm for}\;\;\StoN,
\ee
whose value
fills a gap between the experimental result and theoretical estimate of 
$|V_{us}f_1(0)|_{\StoN}$. 

The prediction of $g_2(0)$ given when combining the experimental information with our result of $f_1(0)$ 
is roughly consistent with the size of the first order corrections and in agreement with the numerical results
of the $g_2(q^2)$ form factor directly calculated 
in quenched lattice QCD~\cite{{Guadagnoli:2006gj},{Sasaki:2008ha}}.
Our preliminary result from 2+1 flavor dynamical lattice QCD has been 
reported in Ref.~\cite{Sasaki:2016zpr} and further study is now in progress~\cite{inprogress}. 
Although it is most likely that the CKM unitarity could be satisfied in the $\StoN$ 
beta decay within
the current experimental accuracy, the confirmation of the non-zero value 
of $g_2(0)$ directly calculated from the first-principles
is primary required for the first-row CKM-unitarity test through 
independent determinations of $V_{us}$ from the hyperon beta decays.

%
%
\begin{figure}[ht]
\centering
\includegraphics[width=.48\textwidth,clip]{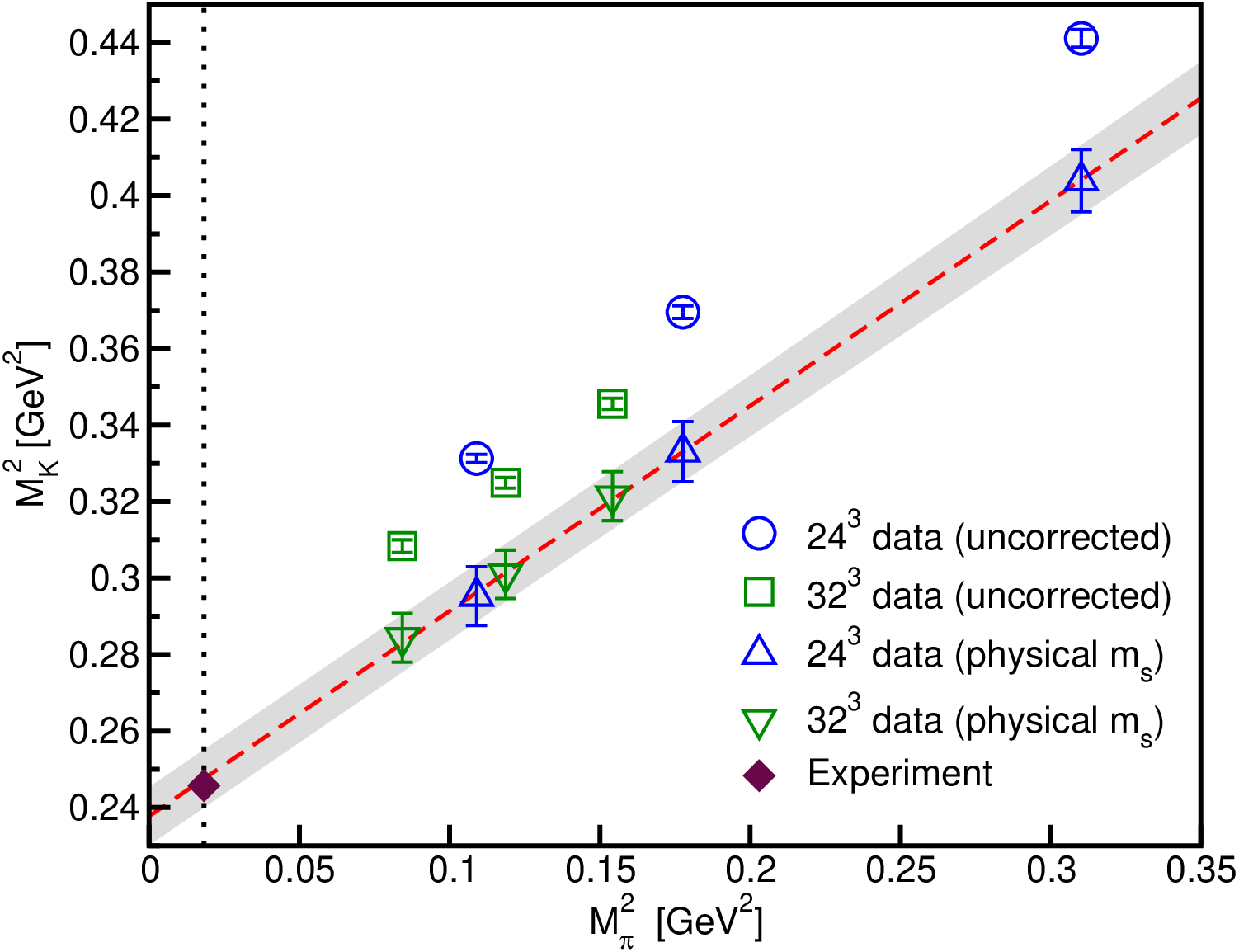}
\caption{Chiral behavior of the kaon mass as a function of the pion mass squared 
in the physical units. Open circle (squared) symbols denote the original data calculated 
on the $24^3$ ($32^3$) ensembles, while open up-triangle (down-triangle) symbols are
corrected ones using Eq.(\ref{Eq:KaonPhys}), respectively. The dashed line represent
the linear fit on the corrected data. The filled diamond symbol denotes 
the experimental point ($M_\pi=135.0$ MeV and $M_K=495.7$ MeV).
}\label{Fig:GMOR_raw_sft}
\end{figure}
%

%
%
\begin{table}[ht]
\caption{
Results for the kaon mass and the continuum value of $\tilde{f}_1(0)=f_1(0)/f_1^{\rm SU(3)}(0)$ at the physical strange-quark mass. The first error is the statistical uncertainty, while the second error is due to the uncertainty on $m_s^{\rm phys}$. 
}
\label{Tab:f1_SFT}
\begin{ruledtabular}
\begin{tabular}{c l  l  l  l }
\hline
& & \multicolumn{1}{c}{$M_K$ ($m_{s}^{\rm phys}$)}
&  \multicolumn{2}{@{}c@{}}{$\tilde{f}_1(0)$  ($m_{s}^{\rm phys}$)} \cr
$\beta$ &$am_{ud}$
& \multicolumn{1}{c}{[GeV]} 
& \multicolumn{1}{c}{$\StoN$} 
& \multicolumn{1}{c}{$\XtoS$} \cr
\hline
2.13&
0.005 & 0.543(1)(7) &  0.9719(96)(26) & 0.9835(63)(15) \cr
&
0.010   & 0.577(1)(7) &  0.9841(54)(23) & 0.9892(36)(13) \cr
&
0.020  & 0.636(2)(6)& 0.9911(12)(15) & 0.9950(8)(9) \cr
\hline
2.25&
0.004 & 0.533(1)(6) & 0.9672(151)(22) & 0.9816(74)(13)  \cr
&
0.006 & 0.549(1)(6) & 0.9654(73)(21)& 0.9819(42)(12)  \cr
&
0.008 & 0.567(1)(6) & 0.9733(53)(20)& 0.9854(32)(11)  \cr
\hline
\end{tabular}
\end{ruledtabular}
\end{table}

%
%
\begin{figure*}[ht]
\centering
\includegraphics[width=.48\textwidth,clip]{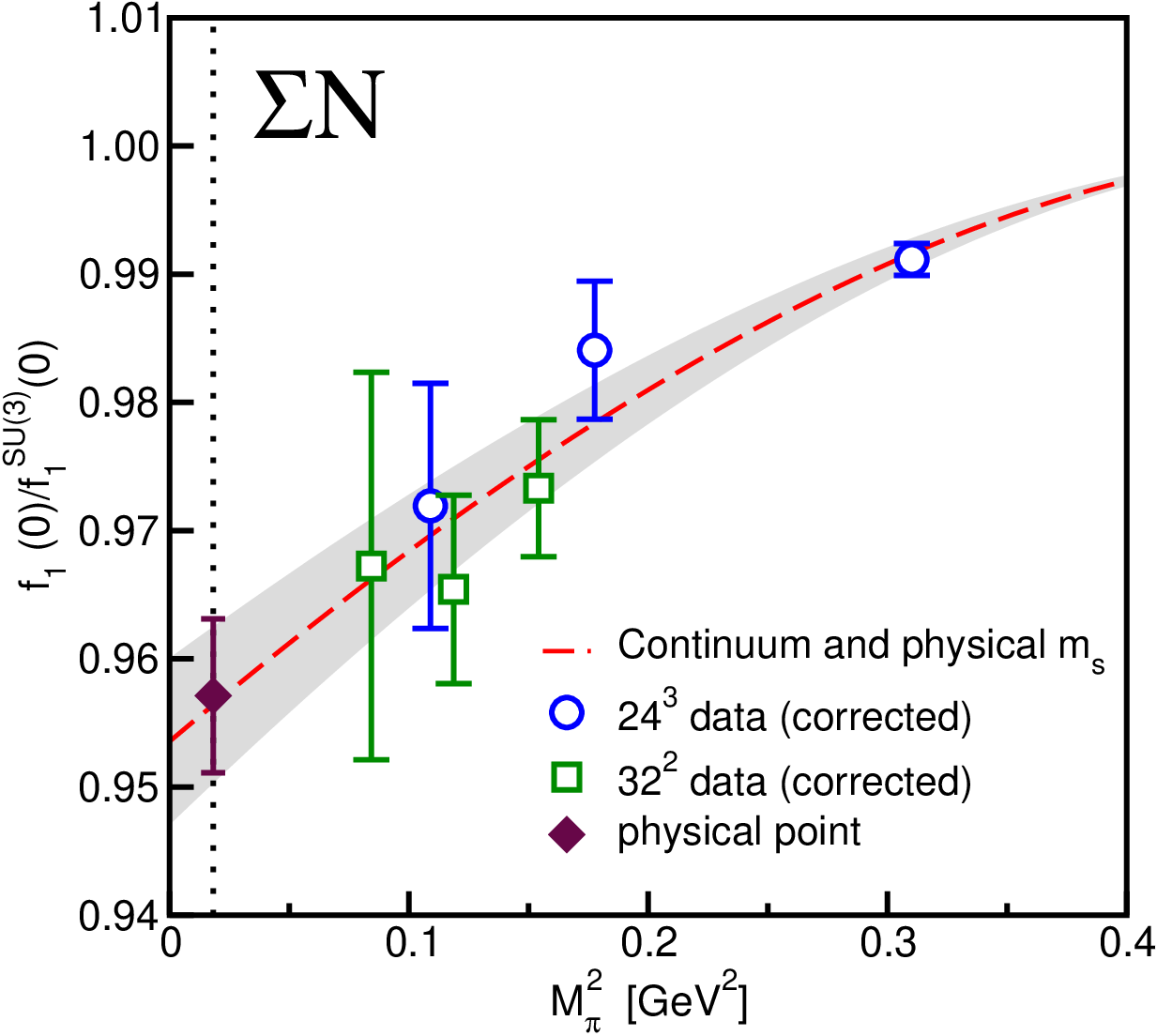}
\includegraphics[width=.48\textwidth,clip]{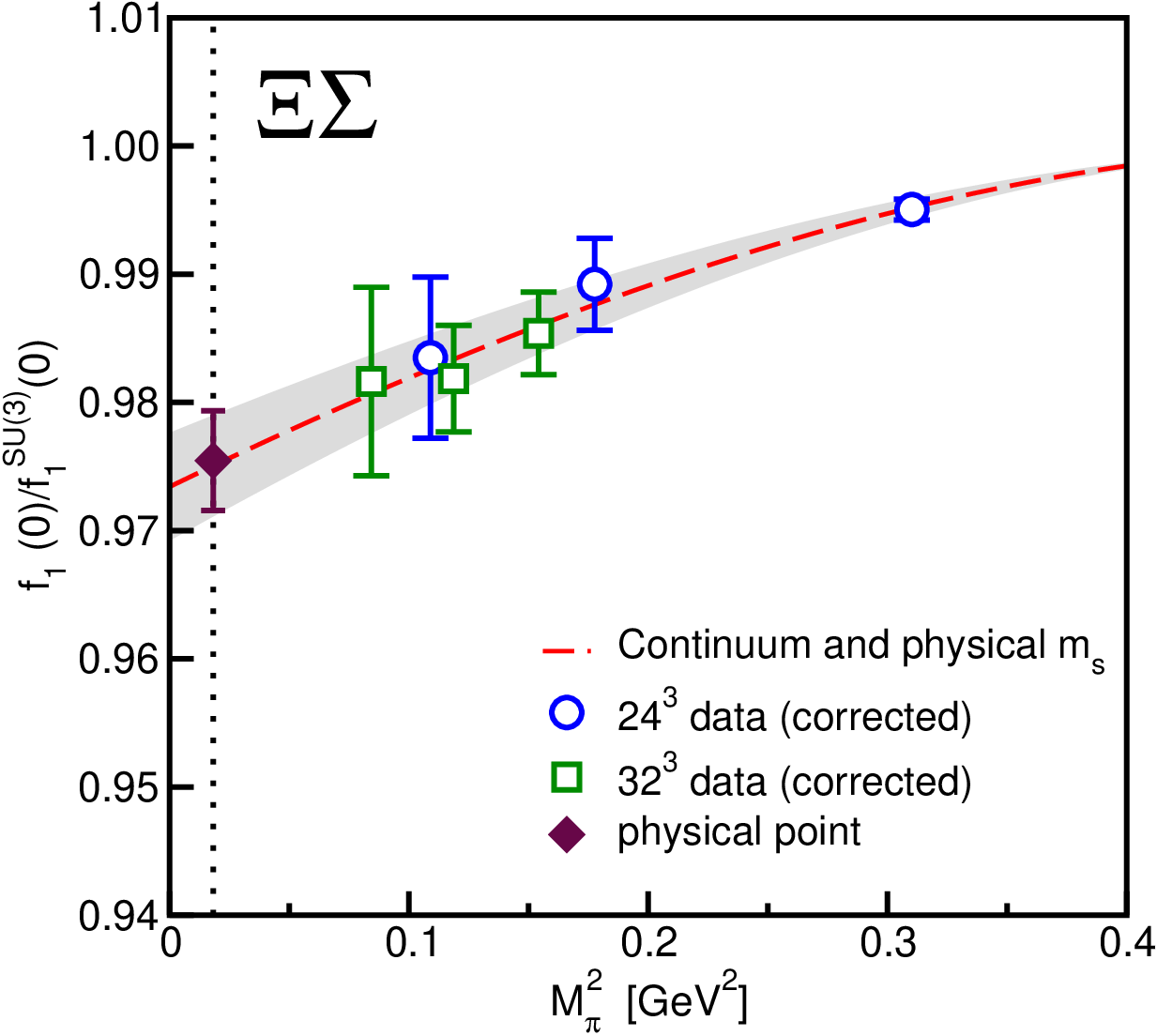}
\caption{
Chiral and continuum extrapolation of $\tilde{f}_1(0)$ for $\StoN$ (left panel) 
and $\XtoS$ (right panel) beta decays. As opposed to Fig.~\ref{Fig:ChiraExtraSft_wasq_su3fix}, 
the data plotted in each panel has been corrected to the continuum limit at the physical strange-quark mass
using the corresponding corrections obtained by the combined 
continuum-chiral fit with Eq.~(\ref{Eq:GlobalFit_type3}) (Type 3 fit).}\label{Fig:ChiraExtraSft_wasq_su3fix}
\end{figure*}
%

\subsection{Evaluation of $f_1(0)$ at the physical strange-quark mass}
\label{Sec:Sec3_corrections}

In Sec.~\ref{Sec:Sec3_chiral_cont}, we have performed the combined chiral-continuum 
extrapolation with all data of $f_1(0)$ calculated at two different lattice spacings 
in order to evaluate results of $f_1(0)$ in the continuum limit and at physical quark masses. 
The functional form (Type 3) of the combined global fit is designed to eliminate the 
leading errors associated with discretization effects and also the 
untuned strange-quark mass corrections. 
The former ${\cal O}(a^2)$ corrections are easily eliminated 
from the data itself with the resulting $C_3$ coefficient.
In order to correct the latter error, we use the following strategy.

In Ref.~\cite{Aoki:2010dy}, the physical strange-quark masses
on both $24^3$ and $32^3$ ensembles have been already determined 
through a reweighting technique as summarized in Table~\ref{Tab:Summary_DWFSim}.
We first evaluate the kaon mass at the physical strange-quark mass ($m_s^{\rm phys}$)
and a given light-quark mass ($m_{ud}$) with a help of the Gell-Mann-Oakes-Renner relation 
for the pion and kaon masses, which correspond to the quark mass dependence of 
pseudo-scalar meson masses at the leading order of ChPT:
%
%
\bea
M_{\pi}^2&=&2B_0 m_{ud}, \\
M_{K}^2&=&B_0(m_{ud}+m_s),
\eea
where $m_s$ represents the simulated strange-quark mass and the constant
parameter $B_0$ is related to the scalar quark condensate. 
At this order, the kaon mass at the physical strange-quark mass can 
be easily evaluated by a simple relation,
%
%
\be
M_K^2(m_s^{\rm phys})=\left(M_K^2(m_s)-\frac{1}{2}M_{\pi}^2\right)\frac{m_{s}^{\rm phys}}{m_{s}}
+\frac{1}{2}M_{\pi}^2.
\label{Eq:KaonPhys}
\ee
In Fig.~\ref{Fig:GMOR_raw_sft}, we plot the kaon mass obtained by Eq.~(\ref{Eq:KaonPhys}) as
a function of $M_{\pi}^2$. Open circle (squared) symbols denote the original data calculated 
on the $24^3$ ($32^3$) ensembles, while open up-triangle (down-triangle) symbols are
corrected ones by using Eq.~(\ref{Eq:KaonPhys}). 

After correcting towards the physical strange-quark mass using the ans\"atz
in Eq.~(\ref{Eq:KaonPhys}), all data points line up on a dashed line, which represents 
the simple linear chiral extrapolation of all corrected data. The filled diamond symbol denotes 
the experimental point ($M_\pi=135.0$ MeV and $M_K=495.7$ MeV).
Figure~\ref{Fig:GMOR_raw_sft} shows that the chiral behavior of the kaon mass squared can be well
approximated by a linear dependence between the simulated range of masses and the physical point.

%
%
\begin{table*}[ht]
\caption{
Results for $R_{\Delta f}$ in units of $({\rm GeV})^{-4}$.
The data tabulated in the third and fifth columns are
the uncorrected data, while the data tabulated in the fourth and sixth columns have been 
corrected to the continuum limit at the physical strange-quark mass using the corresponding corrections 
obtained by the combined continuum-chiral fit with Eq.~(\ref{Eq:GlobalFit_type3}) (Type 3 fit).
The first error is the statistical uncertainty, while the second error is due to the uncertainty on $m_s^{\rm phys}$.
}
\label{Tab:R_DF_ALL}
\begin{ruledtabular}
\begin{tabular}{cl  ll  ll}
\hline
&&  \multicolumn{2}{@{}c@{}}{$\StoN$} & \multicolumn{2}{@{}c@{}}{$\XtoS$}\cr
$\beta$& $m_{ud}$ 
& No corrections  & Continuum  ($m_{s}^{\rm phys}$) &  No corrections& Continuum ($m_{s}^{\rm phys}$)\cr
\hline
2.13&
0.005 
& $-0.581(187)$ & $-0.809(276)(7)$
&$-0.522(120)$ & $-0.475(181)(63)$ 
\cr
&
0.01   
& $-0.409(136)$ & $-0.659(223)(24)$ 
& $-0.521(89)$ & $-0.446(148)(46)$
\cr
&
0.02   
& $-0.255(107)$ & $-1.008(141)(6)$
& $-0.660(81)$ & $-0.564(93)(46)$\cr
\hline
2.25&
0.004 
& $-0.688(295)$ & $-0.818(377)(4)$
& $-0.488(143)$ & $-0.459(184)(33)$ \cr
&
0.006 
& $-0.839(169)$ &$-1.040(221)(10)$
& $-0.562(94)$ & $-0.546(125)(42)$ \cr
&
0.008 
& $-0.747(144)$ & $-0.954(191)(3)$ 
& $-0.549(83)$ & $-0.523(115)(70)$ \cr
\hline
&
\multicolumn{1}{@{}c@{}}{physical point} & \multicolumn{1}{@{}c@{}}{N/A} & $-0.829(116)$  & \multicolumn{1}{@{}c@{}}{N/A} & $-0.474(75)$\cr
\end{tabular}
\end{ruledtabular}
\end{table*}

Using the corrected kaon mass together with the aforementioned chiral-continuum extrapolation, 
we thus can eliminate the untuned strange-quark mass errors 
from our results of $f_1(0)$ obtained with strange quark masses slightly heavier 
than the physical mass. 
In Fig.~\ref{Fig:ChiraExtraSft_wasq_su3fix}, the resulting values of $\tilde{f}_1(0)$ 
in the continuum limit and at the physical strange-quark mass are shown with
the curve obtained from the aforementioned chiral-continuum global fit.  
As opposed to Fig.~\ref{Fig:ChiraExtraRaw_wasq_su3fix},
the data plotted in each panel has been corrected to the continuum limit at the physical
strange-quark mass. We summarize the values of $f_1(0)$ in the continuum limit and 
at the physical strange-quark mass as well as the corrected kaon masses in Table~\ref{Tab:f1_SFT}. 

%
%
\begin{figure*}[ht]
\centering
\includegraphics[width=.48\textwidth,clip]{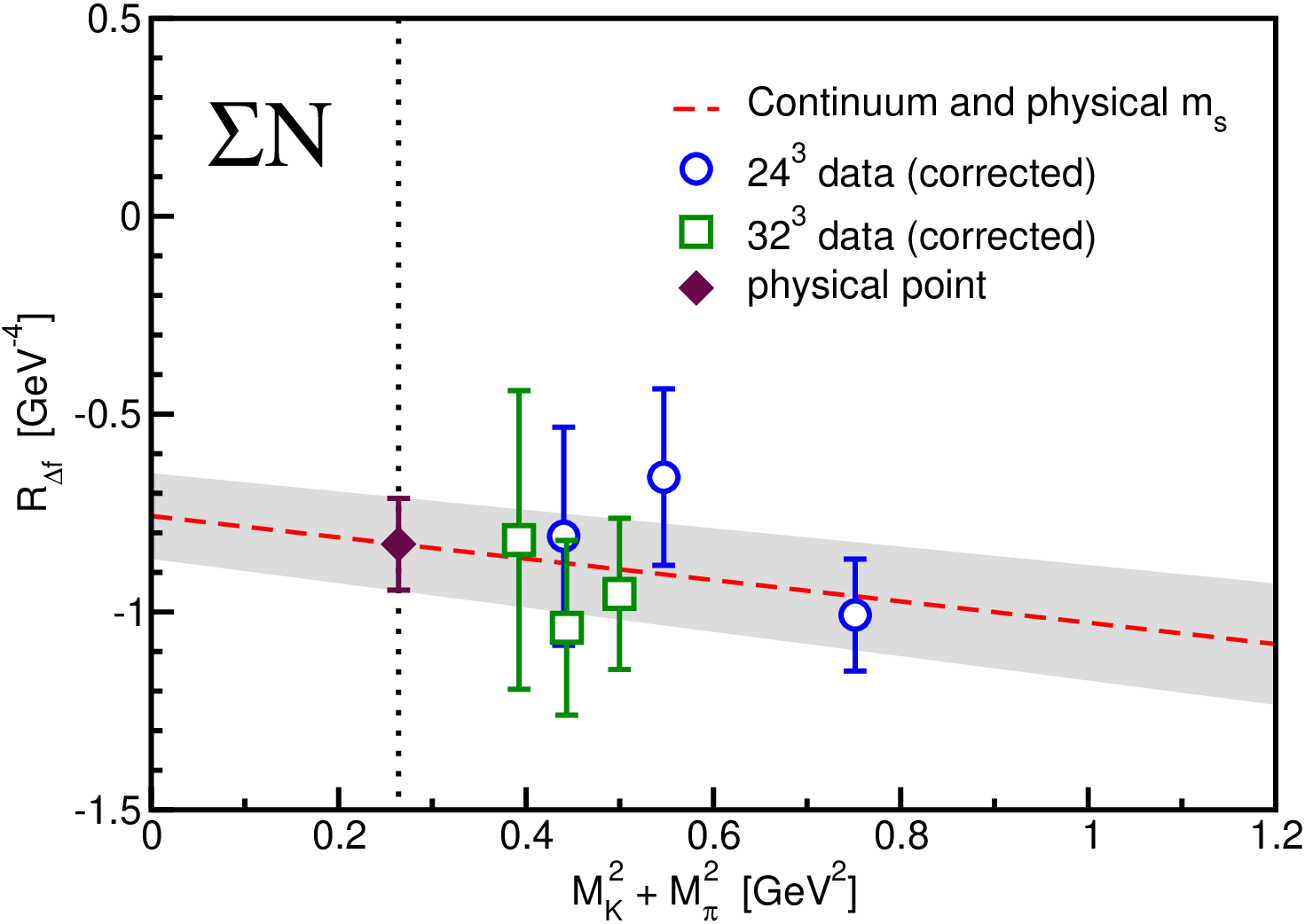}
\includegraphics[width=.48\textwidth,clip]{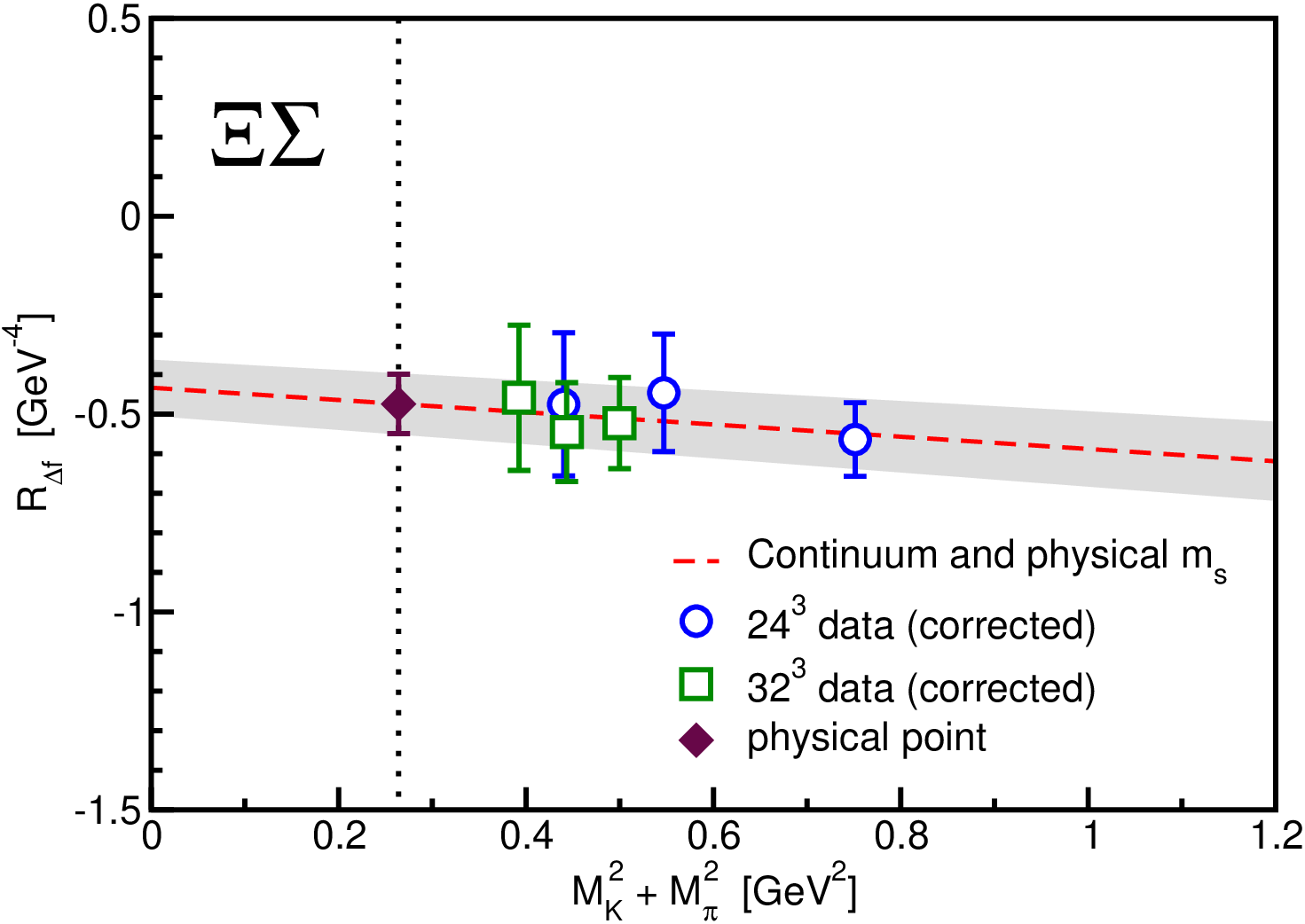}
\caption{Chiral behavior of the continuum values of $R_{\Delta f}$
at the physical strange-quark mass as a function of $M_K^2+M_\pi^2$ in physical units.
The left panel is for the $\StoN$ beta decay, while the right panel is for the $\XtoS$ beta decay. 
In each panel, the dashed curve is obtained by the fit result from the combined 
continuum-chiral extrapolation of the data $\tilde{f}_1(0)$ 
with Eq.~(\ref{Eq:GlobalFit_type3}) (Type 3 fit).
}\label{Fig:ChiraExtra_DRF}
\end{figure*}
We finally evaluate the following ratio:
\be
R_{\Delta f}(M_{K}, M_{\pi}) =\frac{\Delta f}{(M_K^2-M_\pi^2)^2},
\ee
where the leading symmetry-breaking correction, which is predicted 
by the AGT, is explicitly factorized out~\cite{{Guadagnoli:2006gj},{Sasaki:2008ha}}.
In Table~\ref{Tab:R_DF_ALL}, we summarize the values of $R_{\Delta f}$ in the continuum limit and at the physical strange-quark mass as well as those uncorrected values of $R_{\Delta f}$. 
As shown in Fig.~\ref{Fig:ChiraExtra_DRF}, the chiral behavior of the corrected $R_{\Delta f}$, where 
both the discretization effects and the untuned strange-quark mass corrections are eliminated,  
shows neither the higher-order corrections of the $SU(3)$ breaking or the effects of the chiral loops 
predicted by the covariant baryon ChPT~\cite{Geng:2014efa}
in the full range of simulated pion masses. Therefore, our limited data set does not allow to use 
more sophisticated fitting formula of the chiral extrapolation, which is based on the baryon 
ChPT~\footnote{Although, strictly speaking, our simulated pion masses would be 
beyond the range of applicability of the chiral expansion, no large chiral-loop effect found
in $\Delta f$ suggests that the baryon ChPT encounters the convergence problem for $f_1(0)$.}.
In each panel of Fig.~\ref{Fig:ChiraExtra_DRF}, the dashed curve is obtained by the fit result from the combined 
continuum-chiral extrapolation of the data $\tilde{f}_1(0)$ 
with Eq.~(\ref{Eq:GlobalFit_type3}) (Type 3 fit) and the filled diamond symbol corresponds to 
the value of $R_{\Delta f}$ at the physical point. We then quote
these values for both $\StoN$ and $\XtoS$ beta decays:
\be
R_{\Delta f}(M_K^{\rm phys}, M_{\pi}^{\rm phys})=\left\{
\begin{array}{ll}
-0.829(116) & {\rm for}\;\StoN \cr
-0.474(75) & {\rm for}\;\XtoS,
\end{array}
\right.
\ee
which are given in units of $({\rm GeV})^{-4}$.

\section{Summary}
\label{Sec:Sec4}

We have studied the $SU(3)$-breaking effects on the hyperon vector couplings
$f_1(0)$ for the $\StoN$ and $\XtoS$ beta decays with (2+1)-flavors of dynamical 
quarks and calculated $f_1(0)$, for the first time, in the continuum limit. 
Our simulations are carried out with gauge configurations generated by
the RBC and UKQCD Collaborations with (2+1)-flavors of dynamical domain-wall fermions
and the Iwasaki gauge action. Our earlier calculation of $f_1(0)$ was performed on an ensemble
set at a single coarse lattice spacing ($a\approx 0.114$ fm)~\cite{Sasaki:2012ne}. In this 
paper we repeat the calculation at a second value of the finer lattice spacing ($a\approx 0.086$ fm), 
allowing for a continuum extrapolation. 

We first confirm our finding, first presented in Ref.~\cite{Sasaki:2012ne}, that $\Delta f$, which represents 
full $SU(3)$-breaking corrections on $\tilde{f}_1(0)=f_1(0)/f_1^{\rm SU(3)}(0)$, is certainly {\it negative} 
for both beta decays at the finer lattice spacing with the simulated pion mass in the range $M_\pi=290$-393 MeV.
We then performed a combined global-fit of both $24^3$ (coarse) and $32^3$ (fine) lattice data
on $\tilde{f}_1(0)$ to determine the hyperon vector coupling in the continuum limit at the physical point. 
The continuum values of $\tilde{f}_1(0)$ at the physical point reach an accuracy of about 1\% (or less)
and the full $SU(3)$-breaking corrections are estimated to be 4.3\% (2.5\%) for the $\StoN$ ($\XtoS$) 
beta decay. The results are presented in Eq.(\ref{Eq:FinalResult_f1}) and Eq.(\ref{Eq:Delta_f}).

The theoretical estimate of the hyperon vector coupling $f_1(0)$ reaches a sub percent level accuracy.
We thus found that the current $\StoN$ data with lattice input of $f_1(0)$ moves slightly
off the CKM unitarity condition. Conversely, we deduce that this observation would expose a size of the induced second-class form factor $g_2$, 
which was less-known and ignored in experiments~\cite{Cabibbo:2003cu}. 
Indeed, under the assumption of the CKM unitarity, we can estimate $g_2(0)= 0.57(20)$ 
for the $\StoN$ beta decay, whose value fills a gap between the experimental
result and theoretical estimate of $|V_{us}f_1(0)|_{\StoN}$.

Our prediction of $g_2(0)$ is roughly consistent with 
the size of the first-order $SU(3)$ symmetry-breaking corrections and also in 
agreement with the results of the $g_2(q^2)$ form factor directly calculated 
in quenched lattice QCD~\cite{{Guadagnoli:2006gj},{Sasaki:2008ha}}.
Thus, it is most likely that the CKM unitarity could be satisfied in the $\StoN$ beta decay 
within the current experimental accuracy.

The confirmation of the non-zero value 
of $g_2(0)$ directly calculated from the first-principles is primary required for independent 
determinations of the CKM matrix element $V_{us}$ from the hyperon beta decays.
In our preliminary calculation, which is reported in Ref.~\cite{Sasaki:2016zpr}, 
a non-zero $g_2$ form factor is likely evident in fully dynamical lattice QCD 
and its size is roughly consistent with the indirect estimation presented here. 
Further study is now in progress~\cite{inprogress}.

\begin{acknowledgments}
It is a pleasure to acknowledge the technical help of P. Boyle and C. Jung for numerical 
calculations on the IBM BlueGene/Q supercomputer.
This work is supported by the Large Scale Simulation Program 
(No.12/13-02, No.13/14-03, No.14/15-02, No.15/16-01, No.16/17-01) 
of the High Energy Accelerator Research Organization (KEK) and also the 
Interdisciplinary Computational Science Program (13a-1, 14a-1, 15a-1, 16a-1)
in the Center for Computational Sciences, University of Tsukuba.
Numerical calculations reported here were performed (in part) using the KEK supercomputer system,
the COMA (PACS-IX) system at the CCS, University of Tsukuba, and also the RIKEN Integrated Cluster
of Clusters (RICC) facility.
\end{acknowledgments}


\end{document}